\documentclass[zpreprint,zbstepj,british]{zeus_K2.9paper}
\usepackage{graphicx}
\usepackage{booktabs}
\usepackage{multirow}
\usepackage{comment}
\usepackage{caption}
\usepackage{subcaption}
\usepackage{lineno}
\usepackage{stackengine}

\usepackage{rotating}
\usepackage{amsmath}
\usepackage{verbatim}
\usepackage{stackengine}
\usepackage{pdfpages}
\usepackage{longtable}
\usepackage{ulem}

\usepackage{zeusdet_hepunits}
\usepackage{array, tabularx, booktabs, longtable}
\def\citeCTD{{\cite{%
nim:a279:290,*npps:b32:181,*nim:a338:254%
}}\xspace}
\def\citeMVD{{\cite{%
nim:a581:656%
}}\xspace}
\def\citeSTT{{\cite{%
nim:a535:191%
}}\xspace}

\def\citeHES{{\cite{%
nim:a277:176%
}}\xspace}
\def\citeSixmT{{\cite{%
thesis:gosau:2007%
}}\xspace}
\def\citeBAC{{\cite{%
nim:a300:480%
}}\xspace}
\def\citePCAL{{\cite{%
desy-92-066,*zfp:c63:391,*acpp:b32:2025%
}}\xspace}
\def\citeSPECTRO{{\cite{%
nim:a565:572%
}}\xspace}

\newcommand{\Nrec}{\ensuremath{N_{\textrm{rec}}}\xspace}
\newcommand{\Nch}{\ensuremath{N_{\textrm{ch}}}\xspace}
\newcommand{\cn}[1]{\ensuremath{c_{#1}{\{2\}}}\xspace}
\newcommand{\cnFour}[1]{\ensuremath{c_{#1}\{4\}}\xspace}
\newcommand{\Cdetadphi}{\ensuremath{C(\Delta\eta, \Delta\varphi)}\xspace}
\newcommand{\deta}{\ensuremath{|\Delta \eta|}\xspace}

\newcommand{\pT}{\ensuremath{p_{\rm T}}\xspace}
\newcommand{\pTone}{\ensuremath{p_{\rm T,1}}\xspace}
\newcommand{\meanpt}{\ensuremath{\left< \pT \right>}\xspace}
\newcommand{\ep}{\ensuremath{ep}\xspace}
\newcommand{\pTref}{\ensuremath{p_{\textrm{T}0}^{\rm ref}}\xspace}

\usepackage{array}

\usepackage{dcolumn}
\usepackage{color,soul} 
\usepackage{color}
\newcolumntype{d}[1]{D{.}{.}{#1} }

\begin{document}
%
%
%
\prepnum{DESY--21--099}
\draftversion{}
\prepdate{December 2021}

\title{\Large Azimuthal correlations in photoproduction and deep inelastic $\boldsymbol{ep}$ scattering at HERA}
                  
\author{ZEUS Collaboration}
\date{}              

\maketitle


\begin{abstract}\noindent
{Collective behaviour of final-state hadrons, and multiparton interactions are studied in high-multiplicity $ep$ scattering at a centre-of-mass energy $\sqrt{s}=318 \GeV$ with the ZEUS detector at HERA. 
Two- and four-particle azimuthal correlations, as well as multiplicity, transverse momentum, and pseudorapidity distributions for charged-particle multiplicities $\Nch \geq 20$ are measured.
The dependence of two-particle correlations on the virtuality of the exchanged photon shows a clear transition from  photoproduction to neutral current deep inelastic scattering.
For the multiplicities studied, neither the measurements in photoproduction processes nor those in neutral current deep inelastic scattering indicate significant collective behaviour of the kind observed in high-multiplicity hadronic collisions at RHIC and the LHC. 
Comparisons of PYTHIA predictions with the measurements in photoproduction strongly indicate the presence of multiparton interactions from hadronic fluctuations of the exchanged photon.
}

\end{abstract}

\thispagestyle{empty}
\clearpage

\begin{center}
{                      \Large  The ZEUS Collaboration              }
\end{center}

{\small\raggedright


I.~Abt$^{1}$, 
R. Aggarwal$^{2}$, 
V.~Aushev$^{3}$, 
O.~Behnke$^{4}$, 
A.~Bertolin$^{5}$, 
I.~Bloch$^{6}$, 
I.~Brock$^{7}$, 
N.H.~Brook$^{8, a}$, 
R.~Brugnera$^{9}$, 
A.~Bruni$^{10}$, 
P.J.~Bussey$^{11}$, 
A.~Caldwell$^{1}$, 
C.D.~Catterall$^{12}$, 
J.~Chwastowski$^{13}$, 
J.~Ciborowski$^{14, b}$, 
R.~Ciesielski$^{4, c}$, 
A.M.~Cooper-Sarkar$^{15}$, 
M.~Corradi$^{10, d}$, 
R.K.~Dementiev$^{16}$, 
S.~Dusini$^{5}$, 
J.~Ferrando$^{4}$, 
S.~Floerchinger$^{17}$,
B.~Foster$^{15, e}$, 
E.~Gallo$^{18, f}$, 
D.~Gangadharan$^{19, g}$, 
A.~Garfagnini$^{9}$, 
A.~Geiser$^{4}$, 
L.K.~Gladilin$^{16}$, 
Yu.A.~Golubkov$^{16}$, 
G.~Grzelak$^{14}$, 
C.~Gwenlan$^{15}$, 
D.~Hochman$^{20}$, 
N.Z.~Jomhari$^{4}$, 
I.~Kadenko$^{3}$, 
U.~Karshon$^{20}$, 
P.~Kaur$^{21}$, 
R.~Klanner$^{18}$, 
U.~Klein$^{4, h}$, 
I.A.~Korzhavina$^{16}$, 
N.~Kovalchuk$^{18}$, 
M.~Kuze$^{22}$, 
B.B.~Levchenko$^{16}$, 
A.~Levy$^{23}$, 
B.~L\"ohr$^{4}$, 
E.~Lohrmann$^{18}$, 
A.~Longhin$^{9}$, 
F.~Lorkowski$^{4}$,
O.Yu.~Lukina$^{16}$, 
I.~Makarenko$^{4}$, 
J.~Malka$^{4, i}$, 
S.~Masciocchi$^{24, j}$, 
K.~Nagano$^{25}$, 
J.D.~Nam$^{26}$, 
J.~Onderwaater$^{19, k}$, 
Yu.~Onishchuk$^{3}$, 
E.~Paul$^{7}$, 
I.~Pidhurskyi$^{27}$, 
A.~Polini$^{10}$, 
M.~Przybycie\'n$^{28}$, 
A.~Quintero$^{26}$, 
M.~Ruspa$^{29}$, 
U.~Schneekloth$^{4}$, 
T.~Sch\"orner-Sadenius$^{4}$, 
I.~Selyuzhenkov$^{24}$, 
M.~Shchedrolosiev$^{4}$, 
L.M.~Shcheglova$^{16}$, 
I.O.~Skillicorn$^{11}$, 
W.~S{\l}omi\'nski$^{30}$, 
A.~Solano$^{31}$, 
L.~Stanco$^{5}$, 
N.~Stefaniuk$^{4}$, 
B.~Surrow$^{26}$, 
K.~Tokushuku$^{25}$, 
O.~Turkot$^{4, i}$, 
T.~Tymieniecka$^{32}$, 
A.~Verbytskyi$^{1}$, 
W.A.T.~Wan Abdullah$^{33}$, 
K.~Wichmann$^{4}$, 
M.~Wing$^{8, l}$, 
S.~Yamada$^{25}$, 
Y.~Yamazaki$^{34}$, 
A.F.~\.Zarnecki$^{14}$, 
O.~Zenaiev$^{4, m}$ 
\newpage


{\setlength{\parskip}{0.4em}
\makebox[3ex]{$^{1}$}
\begin{minipage}[t]{14cm}
{\it Max-Planck-Institut f\"ur Physik, M\"unchen, Germany}

\end{minipage}

\makebox[3ex]{$^{2}$}
\begin{minipage}[t]{14cm}
{\it DST-Inspire Faculty, Department of Technology, SPPU, India}

\end{minipage}

\makebox[3ex]{$^{3}$}
\begin{minipage}[t]{14cm}
{\it Department of Nuclear Physics, National Taras Shevchenko University of Kyiv, Kyiv, Ukraine}

\end{minipage}

\makebox[3ex]{$^{4}$}
\begin{minipage}[t]{14cm}
{\it Deutsches Elektronen-Synchrotron DESY, Hamburg, Germany}

\end{minipage}

\makebox[3ex]{$^{5}$}
\begin{minipage}[t]{14cm}
{\it INFN Padova, Padova, Italy}~$^{A}$

\end{minipage}

\makebox[3ex]{$^{6}$}
\begin{minipage}[t]{14cm}
{\it Deutsches Elektronen-Synchrotron DESY, Zeuthen, Germany}

\end{minipage}

\makebox[3ex]{$^{7}$}
\begin{minipage}[t]{14cm}
{\it Physikalisches Institut der Universit\"at Bonn,
Bonn, Germany}~$^{B}$

\end{minipage}

\makebox[3ex]{$^{8}$}
\begin{minipage}[t]{14cm}
{\it Physics and Astronomy Department, University College London,
London, United Kingdom}~$^{C}$

\end{minipage}

\makebox[3ex]{$^{9}$}
\begin{minipage}[t]{14cm}
{\it Dipartimento di Fisica e Astronomia dell' Universit\`a and INFN,
Padova, Italy}~$^{A}$

\end{minipage}

\makebox[3ex]{$^{10}$}
\begin{minipage}[t]{14cm}
{\it INFN Bologna, Bologna, Italy}~$^{A}$

\end{minipage}

\makebox[3ex]{$^{11}$}
\begin{minipage}[t]{14cm}
{\it School of Physics and Astronomy, University of Glasgow,
Glasgow, United Kingdom}~$^{C}$

\end{minipage}

\makebox[3ex]{$^{12}$}
\begin{minipage}[t]{14cm}
{\it Department of Physics, York University, Ontario, Canada M3J 1P3}~$^{D}$

\end{minipage}

\makebox[3ex]{$^{13}$}
\begin{minipage}[t]{14cm}
{\it The Henryk Niewodniczanski Institute of Nuclear Physics, Polish Academy of \\
Sciences, Krakow, Poland}

\end{minipage}

\makebox[3ex]{$^{14}$}
\begin{minipage}[t]{14cm}
{\it Faculty of Physics, University of Warsaw, Warsaw, Poland}

\end{minipage}

\makebox[3ex]{$^{15}$}
\begin{minipage}[t]{14cm}
{\it Department of Physics, University of Oxford,
Oxford, United Kingdom}~$^{C}$

\end{minipage}

\makebox[3ex]{$^{16}$}
\begin{minipage}[t]{14cm}
{\it Lomonosov Moscow State University, Skobeltsyn Institute of Nuclear Physics,
Moscow, Russia}

\end{minipage}

\makebox[3ex]{$^{17}$}
\begin{minipage}[t]{14cm}
{\it Institute for Theoretical Physics, University of Heidelberg, Heidelberg, Germany~$^{K}$}

\end{minipage}

\makebox[3ex]{$^{18}$}
\begin{minipage}[t]{14cm}
{\it Hamburg University, Institute of Experimental Physics, Hamburg,
Germany}~$^{E}$

\end{minipage}

\makebox[3ex]{$^{19}$}
\begin{minipage}[t]{14cm}
{\it Physikalisches Institut of the University of Heidelberg, Heidelberg, Germany~$^{K}$}

\end{minipage}

\makebox[3ex]{$^{20}$}
\begin{minipage}[t]{14cm}
{\it Department of Particle Physics and Astrophysics, Weizmann
Institute, Rehovot, Israel}

\end{minipage}

\makebox[3ex]{$^{21}$}
\begin{minipage}[t]{14cm}
{\it Sant Longowal Institute of Engineering and Technology, Longowal, Punjab, India}

\end{minipage}

\makebox[3ex]{$^{22}$}
\begin{minipage}[t]{14cm}
{\it Department of Physics, Tokyo Institute of Technology,
Tokyo, Japan}~$^{F}$

\end{minipage}

\makebox[3ex]{$^{23}$}
\begin{minipage}[t]{14cm}
{\it Raymond and Beverly Sackler Faculty of Exact Sciences, School of Physics, \\
Tel Aviv University, Tel Aviv, Israel}~$^{G}$

\end{minipage}

\makebox[3ex]{$^{24}$}
\begin{minipage}[t]{14cm}
{\it GSI Helmholtzzentrum f\"{u}r Schwerionenforschung GmbH, Darmstadt, Germany~$^{K}$}

\end{minipage}

\makebox[3ex]{$^{25}$}
\begin{minipage}[t]{14cm}
{\it Institute of Particle and Nuclear Studies, KEK,
Tsukuba, Japan}~$^{F}$

\end{minipage}

\makebox[3ex]{$^{26}$}
\begin{minipage}[t]{14cm}
{\it Department of Physics, Temple University, Philadelphia, PA 19122, USA}~$^{H}$

\end{minipage}

\makebox[3ex]{$^{27}$}
\begin{minipage}[t]{14cm}
{\it Institut f\"ur Kernphysik, Goethe Universit\"at, Frankfurt am Main, Germany}

\end{minipage}

\makebox[3ex]{$^{28}$}
\begin{minipage}[t]{14cm}
{\it AGH University of Science and Technology, Faculty of Physics and Applied Computer
Science, Krakow, Poland}

\end{minipage}

\makebox[3ex]{$^{29}$}
\begin{minipage}[t]{14cm}
{\it Universit\`a del Piemonte Orientale, Novara, and INFN, Torino,
Italy}~$^{A}$

\end{minipage}

\makebox[3ex]{$^{30}$}
\begin{minipage}[t]{14cm}
{\it Department of Physics, Jagellonian University, Krakow, Poland}~$^{I}$

\end{minipage}

\makebox[3ex]{$^{31}$}
\begin{minipage}[t]{14cm}
{\it Universit\`a di Torino and INFN, Torino, Italy}~$^{A}$

\end{minipage}

\makebox[3ex]{$^{32}$}
\begin{minipage}[t]{14cm}
{\it National Centre for Nuclear Research, Warsaw, Poland}

\end{minipage}

\makebox[3ex]{$^{33}$}
\begin{minipage}[t]{14cm}
{\it National Centre for Particle Physics, Universiti Malaya, 50603 Kuala Lumpur, Malaysia}~$^{J}$

\end{minipage}

\makebox[3ex]{$^{34}$}
\begin{minipage}[t]{14cm}
{\it Department of Physics, Kobe University, Kobe, Japan}~$^{F}$

\end{minipage}

}

\vspace{3em}


{\setlength{\parskip}{0.4em}\raggedright
\makebox[3ex]{$^{ A}$}
\begin{minipage}[t]{14cm}
 supported by the Italian National Institute for Nuclear Physics (INFN) \
\end{minipage}

\makebox[3ex]{$^{ B}$}
\begin{minipage}[t]{14cm}
 supported by the German Federal Ministry for Education and Research (BMBF), under
 contract No.\ 05 H09PDF\
\end{minipage}

\makebox[3ex]{$^{ C}$}
\begin{minipage}[t]{14cm}
 supported by the Science and Technology Facilities Council, UK\
\end{minipage}

\makebox[3ex]{$^{ D}$}
\begin{minipage}[t]{14cm}
 supported by the Natural Sciences and Engineering Research Council of Canada (NSERC) \
\end{minipage}

\makebox[3ex]{$^{ E}$}
\begin{minipage}[t]{14cm}
 supported by the German Federal Ministry for Education and Research (BMBF), under
 contract No.\ 05h09GUF, and the SFB 676 of the Deutsche Forschungsgemeinschaft (DFG) \
\end{minipage}

\makebox[3ex]{$^{ F}$}
\begin{minipage}[t]{14cm}
 supported by the Japanese Ministry of Education, Culture, Sports, Science and Technology
 (MEXT) and its grants for Scientific Research\
\end{minipage}

\makebox[3ex]{$^{ G}$}
\begin{minipage}[t]{14cm}
 supported by the Israel Science Foundation\
\end{minipage}

\makebox[3ex]{$^{ H}$}
\begin{minipage}[t]{14cm}
 supported in part by the Office of Nuclear Physics within the U.S.\ DOE Office of Science
\end{minipage}

\makebox[3ex]{$^{ I}$}
\begin{minipage}[t]{14cm}
supported by the Polish National Science Centre (NCN) grant no.\ DEC-2014/13/B/ST2/02486
\end{minipage}

\makebox[3ex]{$^{ J}$}
\begin{minipage}[t]{14cm}
 supported by HIR grant UM.C/625/1/HIR/149 and UMRG grants RU006-2013, RP012A-13AFR and RP012B-13AFR from
 Universiti Malaya, and ERGS grant ER004-2012A from the Ministry of Education, Malaysia\
\end{minipage}

\makebox[3ex]{$^{ K}$}
\begin{minipage}[t]{14cm}
this work is part of and supported by the DFG Collaborative Research 
Centre ``SFB 1225 (ISOQUANT)''\
\end{minipage}
}

\pagebreak[4]
{\setlength{\parskip}{0.4em}


\makebox[3ex]{$^{ a}$}
\begin{minipage}[t]{14cm}
now at University of Bath, United Kingdom\
\end{minipage}

\makebox[3ex]{$^{ b}$}
\begin{minipage}[t]{14cm}
also at Lodz University, Poland\
\end{minipage}

\makebox[3ex]{$^{ c}$}
\begin{minipage}[t]{14cm}
now at Rockefeller University, New York, NY 10065, USA\
\end{minipage}

\makebox[3ex]{$^{ d}$}
\begin{minipage}[t]{14cm}
now at INFN Roma, Italy\
\end{minipage}

\makebox[3ex]{$^{ e}$}
\begin{minipage}[t]{14cm}
also at DESY and University of Hamburg, Hamburg, Germany\
\end{minipage}

\makebox[3ex]{$^{ f}$}
\begin{minipage}[t]{14cm}
also at DESY, Hamburg, Germany\
\end{minipage}

\makebox[3ex]{$^{ g}$}
\begin{minipage}[t]{14cm}
now at University of Houston, Houston, TX 77204, USA\
\end{minipage}

\makebox[3ex]{$^{ h}$}
\begin{minipage}[t]{14cm}
now at University of Liverpool, United Kingdom\
\end{minipage}

\makebox[3ex]{$^{ i}$}
\begin{minipage}[t]{14cm}
now at European X-ray Free-Electron Laser facility GmbH, Hamburg, Germany\
\end{minipage}

\makebox[3ex]{$^{ j}$}
\begin{minipage}[t]{14cm}
also at Physikalisches Institut of the University of Heidelberg, Heidelberg,  Germany\
\end{minipage}

\makebox[3ex]{$^{k}$}
\begin{minipage}[t]{14cm}
now at EUMETSAT, Darmstadt, Germany\
\end{minipage}

\makebox[3ex]{$^{ l}$}
\begin{minipage}[t]{14cm}
also supported by DESY, Hamburg, Germany\
\end{minipage}

\makebox[3ex]{$^{ m}$}
\begin{minipage}[t]{14cm}
now at CERN, Geneva, Switzerland\
\end{minipage}


}
}

\clearpage
\pagenumbering{arabic}

\section{Introduction}
\label{sec-int}

\begin{figure}[!ht]
  \centering
  \begin{subfigure}[b]{0.49\textwidth}
  \includegraphics[width=\textwidth]{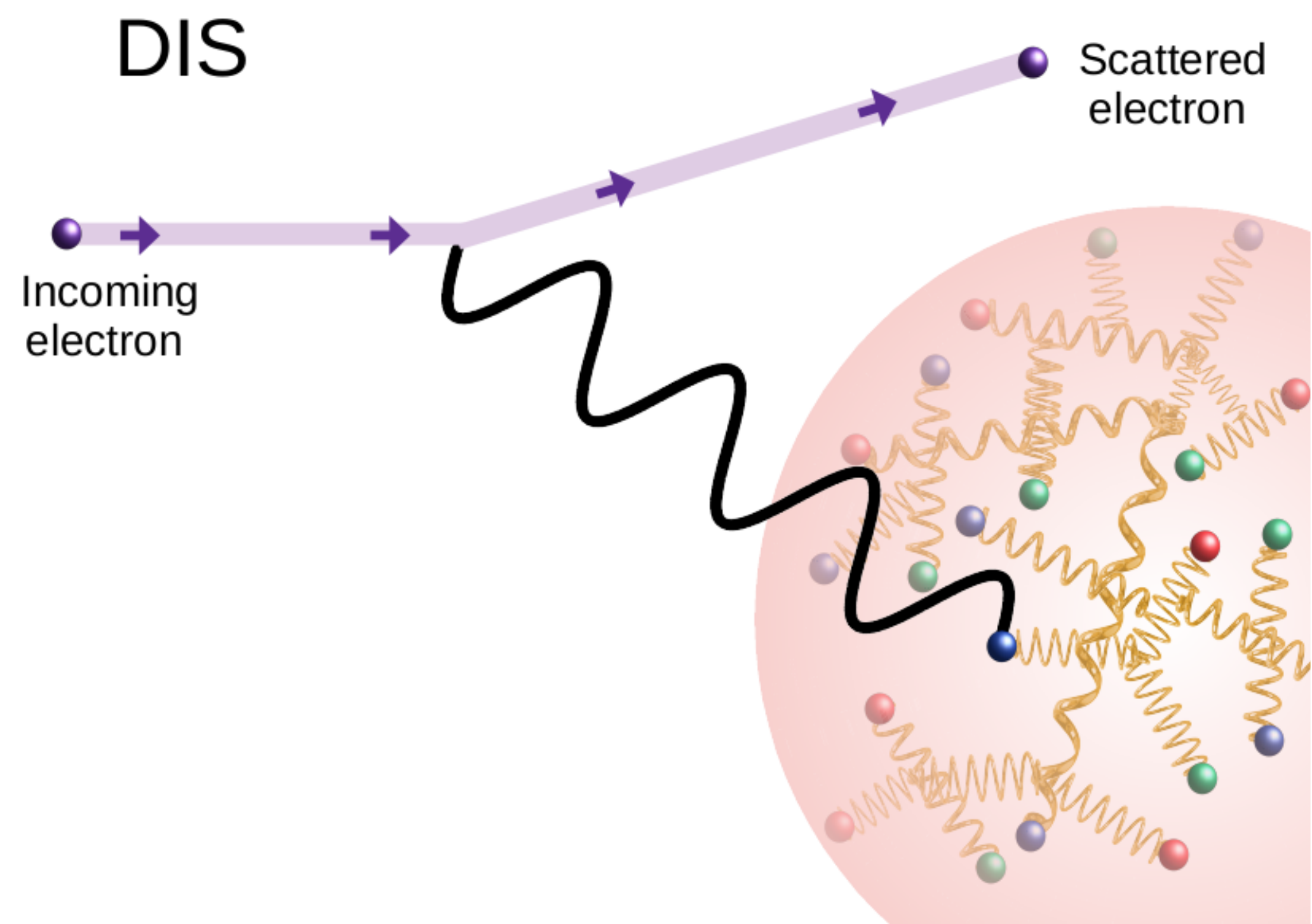}
  \caption{Neutral current deep inelastic scattering.}
  \end{subfigure}
  \begin{subfigure}[b]{0.49\textwidth}
    \includegraphics[width=\textwidth]{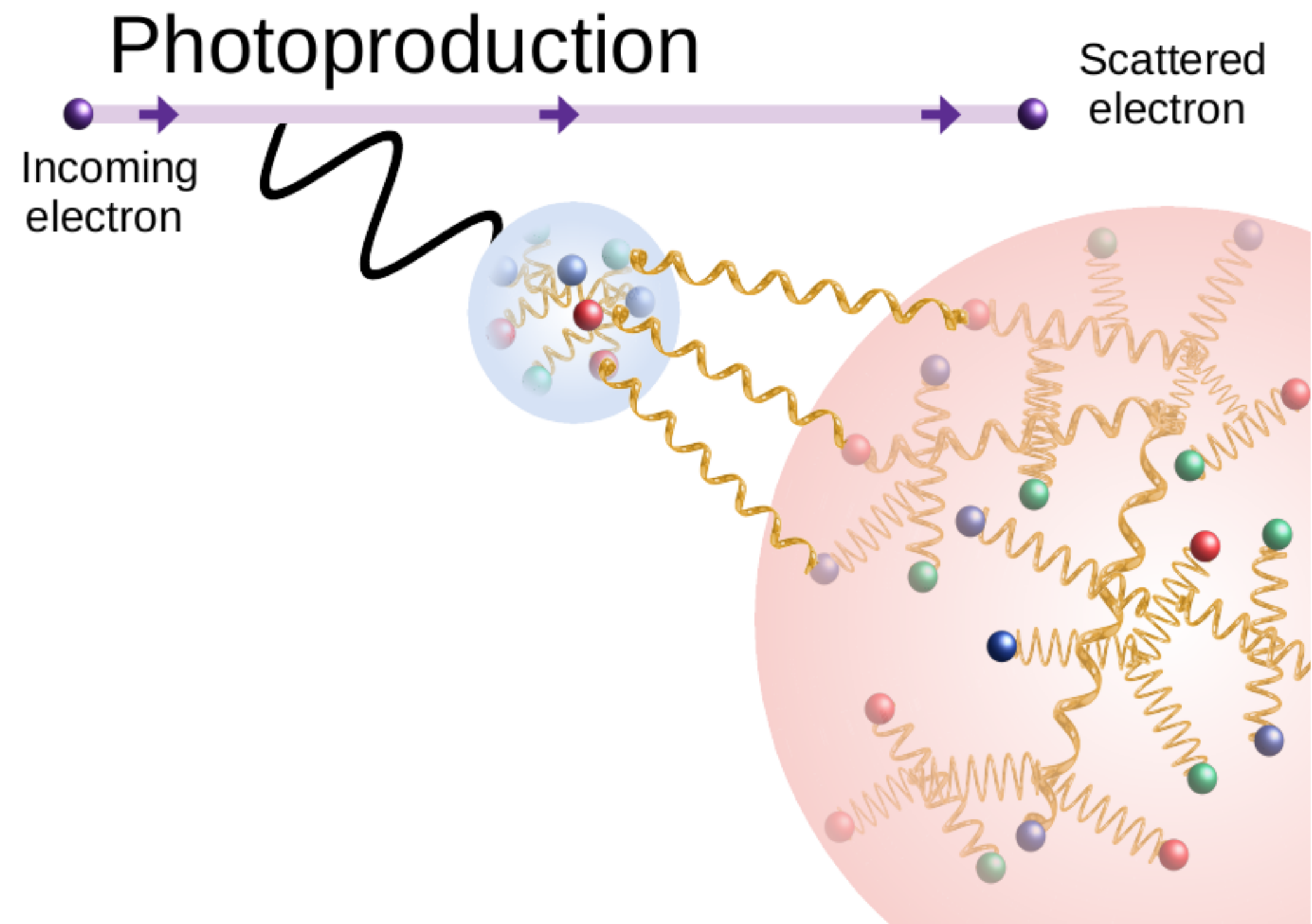}
    \caption{Resolved photoproduction.}
  \end{subfigure}
  \caption{Schematic illustration of initial scattering in (a) deep inelastic scattering and (b) an example of resolved photoproduction.
  The electron beam is represented by the lines with arrows.  The partonic contents of the proton and photon are represented as large and small pale circles, respectively.
  The exchanged photon is shown as a wavy line.  Quarks are shown as spheres while gluons are shown in gold.}
  \label{fig:dis_php_cartoon}
\end{figure}

Two regimes of \ep scattering are distinguished by the virtuality of the
exchanged photon between the electron and proton, which is defined using the square of the four-momentum difference between the incoming and scattered electron as: $Q^2
\equiv -q^2 = -(k-k')^2$.
Neutral current deep inelastic scattering (NC DIS) occurs at large virtualities ($Q^2 \gg 1 \GeV^2$) of the exchanged photon which, at leading order, strikes a single quark within the proton.
Photoproduction ($\gamma p$) processes occur for quasi-real exchanged photons ($Q^2
\lesssim 1 \GeV^2$), and are further sub-divided into two
categories at leading order: direct and resolved.
In direct processes, the photon couples directly to a quark as in DIS.
Resolved processes occur when the photon fluctuates
non-perturbatively into partons, which then scatter with one or more partons in the proton. The DIS and resolved photoproduction regimes are illustrated in Fig.~\ref{fig:dis_php_cartoon}.

A wide variety of measurements in heavy-ion collisions \cite{Afanasiev:2002mx,Arsene:2004fa,Back:2004je,Adams:2005dq,Adcox:2004mh,Aamodt:2010pa} indicates the formation of a new state of quantum chromodynamics (QCD) matter in local thermal equilibrium, the so-called quark-gluon plasma (QGP).
One of the key observables of the QGP is the collective behaviour of final-state particles.
Recent measurements from colliding systems such as $p+p$, $p+A$, and photo-nuclear $A+A$ suggest that a QGP may even form in systems previously thought too small to attain thermal equilibrium \cite{Abelev:2013vea, Khachatryan:2010gv, Abelev:2012ola, Aad:2012gla, Aad:2015gqa, Adare:2013piz, Adare:2015ctn, ATLAS:2021jhn}.
The deep inelastic scattering of leptons on protons produces even smaller systems.
The first search for collective behaviour in such systems was performed using two-particle azimuthal correlations by the ZEUS experiment at HERA \cite{ZEUS:2019jya}.
The measurements demonstrated that the collective effects observed at RHIC and the LHC are not observed in inclusive NC DIS. 

The space-time region probed by the photon in a scattering process can be 
characterised by its de Broglie wavelength and the coherence length of its hadronic fluctuations \cite{Piller:1995kh}.
The resolving power of the exchanged photon increases with its virtuality and is
given by $1/Q$.
Both the coherence length and the wavelength tend to zero for sufficiently large $Q^2$ and in such cases the photon acts as a point-like probe of the proton.
Thus, the probed region in DIS is typically much smaller than the proton
while in photoproduction it can be of order the proton's size, $1/\Lambda_{\textrm{QCD}} \approx 1$ fm.
The characteristics of the interaction region in $\gamma p$ may therefore resemble those in $p+p$ and $p+A$ collisions.

The possibility of observing multiple distinct $2\rightarrow2$ initial partonic
scatterings in a single \ep collision can be investigated with resolved photoproduction at HERA.
Such multiparton interactions (MPI) \cite{Corke:2009tk} have been observed conclusively in high-energy $p+p$ and $p+\Bar{p}$ collisions \cite{Akesson:1986iv, Abe:1997xk, Abazov:2009gc, Aad:2013bjm}.
Indications of MPI have also been observed in \ep photoproduction, such as in analysis of three- and four-jet events \cite{Chekanov:2007ab}.

It is expected that MPI will be plentiful in heavy-ion collisions.
A fully overlapping collision between two lead nuclei, with over 200 nucleons
each, may lead to as many as 1000 binary nucleon collisions \cite{Miller:2007ri}.
Each binary collision may induce further multiple partonic
scatterings, resulting in several thousand MPIs in a single event.
Many measurements in heavy-ion collisions indicate that this dense and extended initial state lays the foundation for a subsequent stage of rescattering between partons, which rapidly come to a local thermal equilibrium \cite{Blok:2017pui}.
The resulting fluid of QCD matter (the QGP) can be described within the framework of relativistic hydrodynamics \cite{Shuryak:1978ij, Bjorken:1982qr, Heinz:2013th}.

\begin{figure}[!ht]
  \centering
  \includegraphics[width=0.5\textwidth]{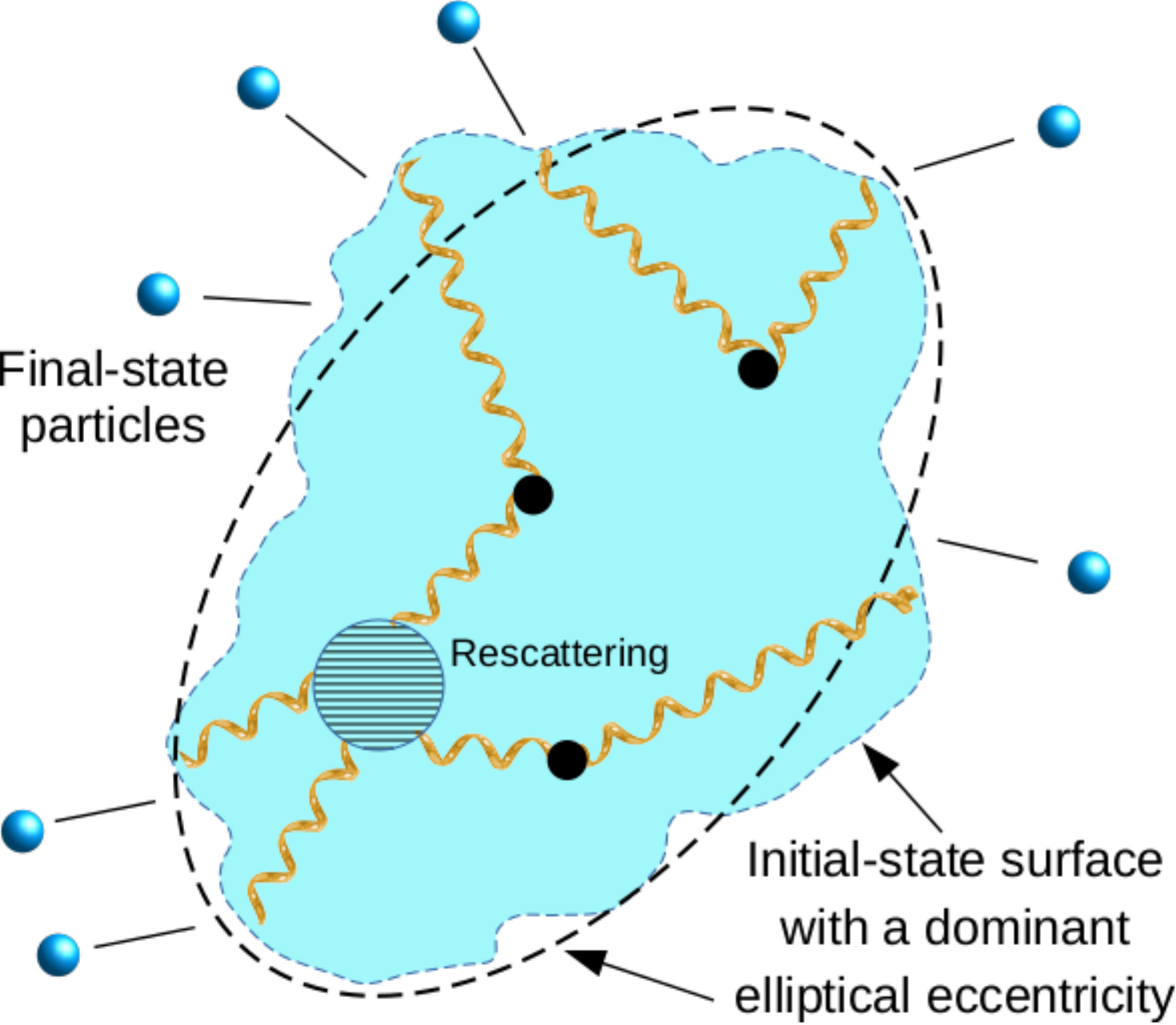}
    \caption{Schematic example of the evolving collision zone in the transverse plane after the initial scattering in resolved photoproduction.  Three MPI centres (shown with solid circles) act as sources of gluons.  The possibility of rescattering of partons from separate MPI centres is illustrated with the hatched circle.  Rescattering is analogous to colour reconnection in PYTHIA.  An initial state with a dominant elliptical eccentricity is shown.  Final-state particles are shown with blue spheres.}
  \label{fig:transverseview_cartoon}
\end{figure}

Photoproduction at HERA provides the opportunity to study MPI and a potential rescattering stage in a larger initial state than NC DIS while still smaller than those in heavy-ion collisions.  
A view of the collision zone in the plane transverse to the beam axis in resolved photoproduction is illustrated in Fig.~\ref{fig:transverseview_cartoon}.
The spatial extent of the zone is in general irregular, but here a zone with a dominant elliptical eccentricity is shown, as often occurs in heavy-ion collisions when two spherical nuclei collide without full overlap in the transverse plane \cite{Drescher:2006pi}.
In $ep$, $p+p$, and $p+A$, elliptical components arise mainly from event-by-event fluctuations of parton distributions \cite{Ma:2014pva}.
In Fig.~\ref{fig:transverseview_cartoon}, three spatially separated MPI centres are depicted as sources of gluons that may further rescatter with other gluons in the system.  
Rescattering is expected to be essential to explain the collective behaviour observed in heavy-ion collisions \cite{Blok:2017pui}.
The possibility of a spatially extended MPI zone and a subsequent rescattering stage in \ep photoproduction thus provides an important intermediate situation between NC DIS, in which no collective behaviour was seen, and larger hadronic systems, where it has been observed.

Azimuthal correlations are sensitive to the dynamics of the collision zone.
Depending on the degree of rescattering, any eccentricities in the initial state can be converted into momentum asymmetries of the produced particles in the final state \cite{Wang:1991qh, Voloshin:1994mz, Borghini:2001vi, Bilandzic:2013kga}.
Two-particle azimuthal correlations can be used to quantify the asymmetries but may be biased by unrelated two-particle correlations such as resonance decays.
Four-particle cumulant correlations are a more robust measure of multiparticle correlations since such two-particle biases are explicitly subtracted off in their construction.

In this article, measurements are presented that are sensitive to collective fluid-like behaviour and MPI in \ep scattering at high charged-particle multiplicity $\Nch \geq 20$.
Additional material for this analysis is provided in the supplementary material.
In photoproduction, measurements are made of the charged-particle multiplicity, pseudorapidity ($\eta = -\ln\left(\tan\frac{\theta}{2}\right)$, where the polar angle, $\theta$, is measured with respect to the $Z$ axis), and transverse momentum (\pT) distributions as well as two- and four-particle azimuthal cumulant correlations.
Two-particle azimuthal correlations are studied as a function of $Q^2$ to illustrate their evolution from photoproduction to DIS.
The possibility of MPI in photoproduction is investigated by comparing the measured distributions and correlation functions to predictions from the PYTHIA 8 event generator \cite{Sjostrand:2014zea}.

\section{Experimental set-up and data selection}
\label{Sec:Experiment}

The photoproduction and NC DIS data used in this analysis were recorded with the ZEUS detector at HERA during 2003--2007 (HERA II).
During this period, the HERA accelerator collided $27.5 \GeV$\ electron/positron beams with $920 \GeV$\ proton beams, which yields a nominal centre-of-mass energy of $\sqrt{s} = 318 \GeV$.
Hereafter, ``electron'' refers to both electrons and positrons unless otherwise stated. 
HERA operated with electron beams during 2005 and part of 2006, while positrons were accelerated in the other years of this data sample.
This analysis uses an integrated luminosity of $366\pm7\,\mathrm{pb}^{-1}$.

\ZcoosysBA \ZpsrapB

A detailed description of the ZEUS detector can be found elsewhere~\cite{zeus:1993:bluebook}.
In the kinematic range of the analysis, charged particles were mainly tracked
in the CTD~\cite{nim:a279:290, npps:b32:181, nim:a338:254} and the microvertex
detector (MVD)~\cite{nim:a581:656}. These detectors operated in a magnetic
field of \unit{1.43}{\tesla} provided by a thin superconducting solenoid.
The high-resolution uranium--scintillator calorimeter (CAL)~\cite{nim:a309:77, nim:a309:101, nim:a321:356, nim:a336:23}
consisted of three parts: the forward (FCAL), the barrel (BCAL) and
the rear (RCAL) calorimeters. Each part was subdivided transversely
into towers and longitudinally into one electromagnetic section (EMC)
and either one (in RCAL) or two (in BCAL and FCAL) hadronic sections
(HAC). \Zlumidesc

The ZEUS experiment operated a three-level trigger system~\cite{uproc:chep:1992:222, Allfrey:2007zz}.
For the NC DIS part of this analysis, 
events were selected at the first level if they had an energy deposit
in the CAL consistent with an isolated scattered electron.
At the second level, a requirement on the total recorded energy and longitudinal momentum of the event was used to select NC DIS event candidates.
At the third level, the full event was reconstructed and tighter requirements for a DIS electron were made.
For the photoproduction analysis, an inclusive set of triggers did not exist.  
Instead, triggers designed to capture heavy-flavour-decay candidates and low-\pT jets were utilised and their corresponding biases to an inclusive measurement were estimated and corrected for using the ZEUS detector Monte Carlo simulation.
The retention of only high-multiplicity events minimised the trigger bias to this analysis, since they are more likely to contain energy deposits above the jet trigger threshold or track combinations resembling heavy-flavour decays.

Photoproduction and NC DIS differ by the absence or presence of a scattered electron in the ZEUS detector.
In photoproduction, the electron typically scatters at small angles and remains undetected, while in NC DIS, the angle of the scattered electron increases with $Q^2$, and above a minimum $Q^2$, can be efficiently detected.
A neural-network algorithm \cite{nim:a365:508,nim:a391:360} was used to identify a scattered electron in the CAL using energy deposits.
In photoproduction, where the scattered electron escapes the detector, the calculated probability and energy of an electron candidate returned from the algorithm are typically small.
Owing to energy and momentum conservation, the difference between the total measured energy and the $Z$-component of momentum in a fully contained event must coincide with twice that of the electron beam: $E-p_Z = 55 \GeV$.
The undetected scattered electron in photoproduction typically leads to values well below this.

Offline event-selection criteria for $\gamma p$ and NC DIS were applied according to the above features.
If a scattered electron was identified by the neural-network algorithm, its assigned probability was required to be less than $90\%$ for $\gamma p$ and greater than $90\%$ for NC DIS.
In addition, the scattered-electron energy in the CAL was required to be less than $15 \GeV$ for $\gamma p$ and larger than $10 \GeV$ for NC DIS. 
The difference of the total observed energy and $Z$-component of momentum, $E-p_Z$, 
was required to be less than $55 \GeV$ for $\gamma p$ and between 47 and $69 \GeV$ for NC DIS.
In NC DIS, the virtuality, $Q^2$, as determined by the electron method \cite{proc:hera:1991:23} was required to be greater than $5 \GeV^2$, where the scattered electron identification probability was large.
The location of the electron-proton scattering, defined as the primary vertex, was required to be near the centre of the detector, $|V_Z | < 30$ cm.
At least $15\%$ of the tracks reconstructed in the event were required to be associated with the primary vertex to reject beam-gas background.

Reconstructed tracks were used in this analysis if their momentum transverse to the beam-axis was $0.1 < \pT < 5.0 \GeV$ and laboratory pseudorapidity $-1.5 < \eta < 2.0$.
The track associated with the scattered electron candidate used to identify the NC DIS event was excluded from the correlation analysis.
Primary charged tracks were selected in the analysis by requiring the distances of closest approach (DCA) to the primary vertex in the transverse and longitudinal directions to be less than 2 cm.
Some secondary tracks, e.g.\ from small-angle scattering in the beam pipe, were retained.
Secondary particles are defined as those which do not originate from the primary vertex.
These tracks inherit characteristics of their corresponding charged primary particles, thereby retaining correlations with other primary particles.

High-multiplicity events were selected by requiring the number of efficiency-corrected charged primary particles in the kinematic acceptance, $\Nch$, to be at least 20.
The NC DIS contamination to the $\gamma p$ sample, and vice-versa, has been estimated to be of the order of $1\%$ from studies of Monte Carlo events. 
A total of 5 (0.2) million $\gamma p$ (NC DIS) events at high multiplicity passed the event-selection criteria.
About $90\%$ of the inclusive photoproduction cross section is captured with the event selection and correction procedure.
A more detailed description of other event- and track-selection criteria can be found in an earlier ZEUS publication on this subject\cite{ZEUS:2019jya}.

\section{Monte Carlo generators}
\label{Sec:MC}

In this analysis, photoproduction in $ep$ scattering is defined as processes with $Q^2 < 1 \GeV^2$, and is modelled within the PYTHIA \cite{Sjostrand:2001yu,Sjostrand:2014zea} Monte Carlo event generator, which has been further developed recently \cite{Helenius:2017aqz}.
Version 6.220 of PYTHIA \cite{Sjostrand:2001yu} was used for the extraction of efficiency corrections and the associated systematic uncertainties, for which large detector simulated event samples were available.  
For the comparison of the photoproduction measurements to known physics mechanisms, version 8.303 was used \cite{Sjostrand:2014zea}.
Both the direct and resolved components of photoproduction, as defined by PYTHIA, were included in the Monte Carlo samples, resulting in the proportion of about 1:100 for $\Nch \geq 20$.
For the DIS part of the analysis, the LEPTO 6.5 \cite{cpc:101:108} and ARIADNE 4.12 \cite{cpc:71:15} Monte Carlo event generators were used to extract efficiency corrections.
The LEPTO predictions were used to compare to the corrected DIS data.
The ZEUS detector was simulated using GEANT 3 \cite{tech:cern-dd-ee-84-1}.
Primary generated particles were defined as charged hadrons with a mean proper lifetime $\tau>1$~cm, which were produced directly or from the decay of a particle with $\tau < 1$~cm.

\subsection{PYTHIA 8}

Predictions from PYTHIA 8 are shown in this article.  
The key parameters utilised to generate photoproduction events are listed in the supplementary material.
The quark and gluon content of the proton is parametrised with the NNPDF2.3 Parton Distribution Function (PDF) at leading order \cite{Ball:2012cx}.
Partonic fluctuations arising from the quasi-real photon are parametrised with the CJKL \cite{Cornet:2002iy} PDF.
Parton scattering between both PDFs in PYTHIA photoproduction is parametrised by the $p_{\textrm{T}0}$ parameter, which regulates the infra-red divergences and adjusts the level of MPI.
The energy dependence of $p_{\textrm{T}0}$ is parametrised as $p_{\textrm{T}0}=\pTref \, (W/7 \, \TeV)^{0.215}$, where $W$ is the centre-of-mass energy of the photon-proton system (given in \TeV), which fluctuates event-by-event \cite{Sjostrand:2014zea, Helenius:2017aqz}.  The mean and RMS of the $W$ distribution in ZEUS $\gamma p$ data for $\Nch \geq 20$ are 239 and $43 \GeV$, respectively (see the supplementary material).
Products of separate MPI subprocesses may further interact through the colour reconnection (CR) framework in PYTHIA.
Colour reconnection across separate MPIs is analogous to rescattering in Fig.~\ref{fig:transverseview_cartoon}.

The PYTHIA 8 predictions with and without MPI are compared to the measurements in photoproduction.
Three different levels of MPI are chosen with $\pTref=2$, 3, and $4 \GeV$.
A previous study using charged-particle differential cross sections $d\sigma/d\pT$ above $2 \GeV$, $d\sigma/d\eta$ for $|\eta |< 1$, and jet distributions, found a preferred value of $3 \GeV$ \cite{Helenius:2017aqz}.
The mean number of MPI per event for $\pTref = 2$, 3, and $4 \GeV$, are 8.3, 3.8, and 2.2, respectively.

\subsection{Efficiency corrections}

The distributions and correlation functions measured in this analysis are affected by non-uniform particle-tracking efficiency.
The ratios of reconstructed to generated particles, pairs, and quadruplets give the respective efficiencies.
They are calculated differentially in $\varphi$, $\eta$, \pT and charge for single particles. 
For pairs and quadruplets, they are calculated differentially in $\left< \eta_i - \left< \eta \right> \right>$, $\left< p_{\textrm{T},i} - \left< p_{\textrm{T}} \right> \right>$, charge combination, and event multiplicity. 
The angled brackets denote an average over the particles in each pair or quadruplet and $i$ is the index of a particle in the pair or quadruplet.
Additionally, pair and quadruplet efficiencies are differentially calculated in the azimuthal quantities $\varphi_1 - \varphi_2$ and $\varphi_1 + \varphi_2 - \varphi_3 - \varphi_4$, respectively. 
Correction factors for single particles, pairs, and quadruplets are given by the inverse of the efficiencies and are labelled $w_{i}^{(1)}$, $w_{ij}^{(2)}$, $w_{ijkl}^{(4)}$, respectively.

\section{Analysis method}
\label{Sec:Analysis}

All distributions and correlation functions in this article are shown for high-multiplicity events, which are defined by a weighted sum over the number of reconstructed tracks (\Nrec) passing the selection criteria: $\Nch = \sum^{\Nrec}_{i} w^{(1)}_{i} \geq 20$.
Corrections for the trigger bias are denoted by a further factor $w_T$, which deviates from unity only in photoproduction since the ZEUS experiment was designed with an inclusive set of triggers for NC DIS but not for photoproduction.

The charged-particle multiplicity distribution, $dN/d\Nch$, was measured in photoproduction.
Tracking-efficiency corrections were performed using an unfolding procedure.  
The RooUnfold \cite{Prosper:2011zz} Bayesian algorithm was used with the response matrix obtained from Monte Carlo simulation of the ZEUS detector.
Transverse momentum and pseudorapidity distributions, $dN/d\pT$ and $dN/d\eta$, were also measured and corrected for tracking inefficiencies using the $w^{(1)}$ weights. 
Trigger-bias correction factors for both $dN/d\Nch$ and $dN/d\pT$ ranged between approximately 2 and 1.1 for their low and high ends, respectively.

Double-differential two-particle correlations as a function of $\Delta \eta=\eta_1 - \eta_2$ and $\Delta\varphi = \varphi_1 - \varphi_2$ are defined as:
\begin{equation}
    C(\Delta\eta,\Delta\varphi) = \frac{ S(\Delta\eta,\Delta\varphi) }{ B(\Delta\eta,\Delta\varphi) },
\end{equation}
which are measured in small intervals of $\Delta\eta$ and $\Delta\varphi$.
The number of pairs for the signal and background distributions are given by \mbox{$S(\Delta\eta,\Delta\varphi) = N^{\textrm{same}}_{\textrm{pairs}}(\Delta\eta,  \Delta\varphi)$} and $B(\Delta\eta,\Delta\varphi) = N^{\textrm{mixed}}_{\textrm{pairs}} (\Delta\eta,  \Delta\varphi)$, respectively.
The background distribution represents the component of $S$ arising purely from combinatorics.
These pair distributions were formed by taking the first particle from a given event and the other from either the same event or different events (mixed) with similar values of \Nrec and vertex $Z$ position. 
The $S$ distribution was corrected with $w_T \, w^{(2)}_{ij}$, while $B$ was corrected with $w^{(1)}_i w^{(1)}_j$.
Both distributions were symmetrised along $\Delta\eta$ and then individually normalised to unity.

The two- and four-particle azimuthal correlation functions are defined by
\begin{eqnarray}
    C_{n}\{2\} &\equiv& \left< \cos{[n(\varphi_1 - \varphi_2)]} \right>, \label{eq:cn2Def} \\
    C_{n}\{4\} &\equiv& \left< \cos{[n(\varphi_1 + \varphi_2 - \varphi_3 - \varphi_4)]} \right>, \label{eq:Cn4Def}
\end{eqnarray}
where the azimuthal angle of particle one is given by $\varphi_1$, etc.
The first and second harmonics ($n=1, 2$) are studied in this article.
The angled brackets in Eqs.~(\ref{eq:cn2Def}) and (\ref{eq:Cn4Def}) denote an average over pairs and quadruplets, respectively.
The two-particle cumulant coincides with the two-particle correlation function: $\cn{n}=C_{n}\{2\}$.
Two-particle correlations are measured as functions of several kinematic variables, which are denoted by $\alpha$ in the following. 

The measurements of two- and four-particle azimuthal correlations require acceptance corrections for which the averaging procedure is given by 
{
\begin{eqnarray}
\cn{n}(\alpha) &=& \frac{ w_T(\alpha) \sum\limits_{e}^{N_{\textrm{ev}}} \sum\limits_{\textrm{pairs}(\alpha,e)} [\cos{ [ n (\varphi_{i} - \varphi_{j}) ] } ] }{ \sum\limits_{e}^{N_{\textrm{ev}}} \sum\limits_{\textrm{pairs}(\alpha,e)} [1]}, \label{eq:cn2Diff} \\
    C_{n}\{4\}(\pTone) &=& \frac{ w_T(\pTone) \sum\limits_{e}^{N_{\textrm{ev}}} \sum\limits_{\textrm{quads}(\pTone,e)} [ \cos{ [ n (\varphi_{i} + \varphi_{j} - \varphi_{k} - \varphi_{l}) ] } ] }{ \sum\limits_e^{N_{\textrm{ev}}} \sum\limits_{\textrm{quads}(\pTone,e)} [1]}. \label{eq:Cn4Diff} 
    \end{eqnarray}
}The summation operators, with a general argument $b$, are given by
    \begin{eqnarray}
    \sum_{\textrm{pairs}(\alpha,e)} [ b ] &\equiv& \sum_{i \neq j}^{\Nrec} w_{e} \, w^{(2)}_{ij} \Theta(\alpha - \textrm{low})\Theta(\textrm{high} - \alpha) [ b ], \\
    \sum_{\textrm{quads}(\pTone,e)} [ b ] &\equiv& \sum_{i \neq j \neq k \neq l}^{\Nrec} w_e \, w^{(4)}_{ijkl} \Theta(\pTone-\textrm{low})\Theta(\textrm{high} - \pTone) [ b ], \\
    b &\in& \{1, \cos{[n(\varphi_i - \varphi_j)]}, \cos{[n(\varphi_i + \varphi_j - \varphi_k - \varphi_l)]} \}.
\end{eqnarray}
Two-particle correlations are presented for $\alpha \in \{Q^2, \Delta\eta, \meanpt \}$, where $\meanpt = (p_{\textrm{T,i}} + p_{\textrm{T,j}})/2$.
The first sum over $e$ was performed for all events, $N_{\textrm{ev}}$.
The sums over pairs and quadruplets run over all combinations of selected charged particles that lie within the low and high limits of a particular interval of the variable $\alpha$, in events with multiplicity \Nrec.
Step functions, which are zero (unity) for negative (positive) arguments, are denoted by $\Theta$.
The arguments of two- and four-particle correlations will be shown on the horizontal axes of their respective figures. 
Event weights are denoted by $w_e$. 
Trigger-bias correction factors, $w_T$, in photoproduction were approximately 1.3 for the correlation functions.

The four-particle cumulant is given by
\begin{equation}
    \cnFour{n}(\pTone) \equiv C_{n}\{4\}( \pTone ) - 2 \; c_{n}\{2\}( \pTone ) \times c_{n}\{2\}, \label{eq:cn4Diff}
\end{equation}
where two-particle correlations are explicitly subtracted off.
They are measured as a function of \pTone, the \pT of the first particle, $i$, in Eqs.~(\ref{eq:cn2Diff}) and (\ref{eq:Cn4Diff}).
In Eq.~\ref{eq:cn4Diff}, \cn{n} without arguments represents an integrated quantity, which corresponds to setting the low and high limits of the pair-averaging procedure to their kinematic limits.
Event weights deviated from unity only in the construction of four-particle cumulants.
To reduce the known bias \cite{Bilandzic:2010jr} to the four-particle cumulant caused by wide multiplicity bins, $w_e$ is set to the number of pair or quadruplet combinations in Eqs.~(\ref{eq:cn2Diff}) and (\ref{eq:Cn4Diff}) for the construction of $\cnFour{n}$.

An alternative approach to calculating \cn{n} is to decompose
$C(\Delta\eta,\Delta\varphi)$ into Fourier series with coefficients
denoted as $V_{n\Delta}$ or $V_{n,2}$~\cite{Aamodt:2011by}.
In studies of the flow coefficients $v_n$~\cite{Voloshin:2008dg}, which characterise the
anisotropic hydrodynamic expansion in heavy-ion collisions, the \cn{n}
and two-particle flow cumulant $v_n\{2\}$ are related as $v_n\{2\} =
[\cn{n}]^{1/2}$.
The relation between the four-particle flow cumulant $v_n\{4\}$ and
$\cnFour{n}$  is more complicated and is given by $v_n\{4\}=
[-\cnFour{n}]^{1/4}$ for $\pT$ integrated results and $v_n\{4\}(\pT) =
[-\cnFour{n}(\pT)] / [-\cnFour{n}]^{3/4}$ for $\pT$ differential
analysis~\cite{Borghini:2001vi,Bilandzic:2010jr}.
For the two (four)-particle cumulant relation to be applicable, the sign of \cn{n}
($\cnFour{n}$) should be positive (negative).

\section{Systematic uncertainties}
\label{Sec:Systematics}

Systematic uncertainties were estimated by comparing the distributions or correlations obtained with the default event- and track-selection criteria to those obtained with varied settings.
The difference between the results obtained with the default and the varied settings was assigned as a signed systematic uncertainty.
Positive and negative systematic uncertainties were separately summed in quadrature to obtain the total systematic uncertainty.
A full description of the systematic studies performed for the NC DIS part of the analysis can be found in a related ZEUS analysis \cite{ZEUS:2019jya}.
Variations of the track DCA, primary-vertex position, low-\pT tracking efficiency, and different data-taking conditions were done identically for both NC DIS and $\gamma p$.

Additional systematic studies were performed for photoproduction (with values of the uncertainty given for \cn{1} at low $\Delta \eta$ indicated with parenthesis).
The available Monte Carlo photoproduction sample used to extract tracking-efficiency and trigger-bias corrections was found to be biased by a loose jet preselection requirement at generator level, which removed about $10\%$ of high-multiplicity events.
Another Monte Carlo data sample with much stricter jet preselections was utilised to estimate the corresponding bias (about $+5\%$, symmetrised).
The uncertainty from the trigger-bias correction was estimated by comparing the results obtained using three different sets of third-level triggers (about $-25\%$, symmetrised).
After the application of tracking-efficiency corrections in Monte Carlo data, a residual difference remained between the reconstructed and generator-level distributions, as well as reconstructed and generator-level correlations, both caused by the limited dimensionality of the corrections that are generally multidimensional (about $-5\%$).
Offline requirements used to remove NC DIS events from the photoproduction sample were loosened from their default value to the value in parenthesis: $P_e < 0.9 \; (0.98)$, $E_e < 15 \; (30) \GeV$, $E-p_z < 55 \; (65) \GeV$ (about $+5\%$).

\section{Results}
\label{Sec:Results}

Results are presented for charged particles $\Nch \geq 20$ in the kinematic region defined by: $-1.5 < \eta < 2.0$ and $0.1 < \pT < 5.0 \GeV$.
Neutral current DIS is shown for $Q^2 > 5 \GeV^2$ and photoproduction is shown for $Q^2 < 1 \GeV^2$.

\begin{figure}[!ht]
  \centering
  \begin{subfigure}[b]{0.49\textwidth}
  \includegraphics[width=\textwidth]{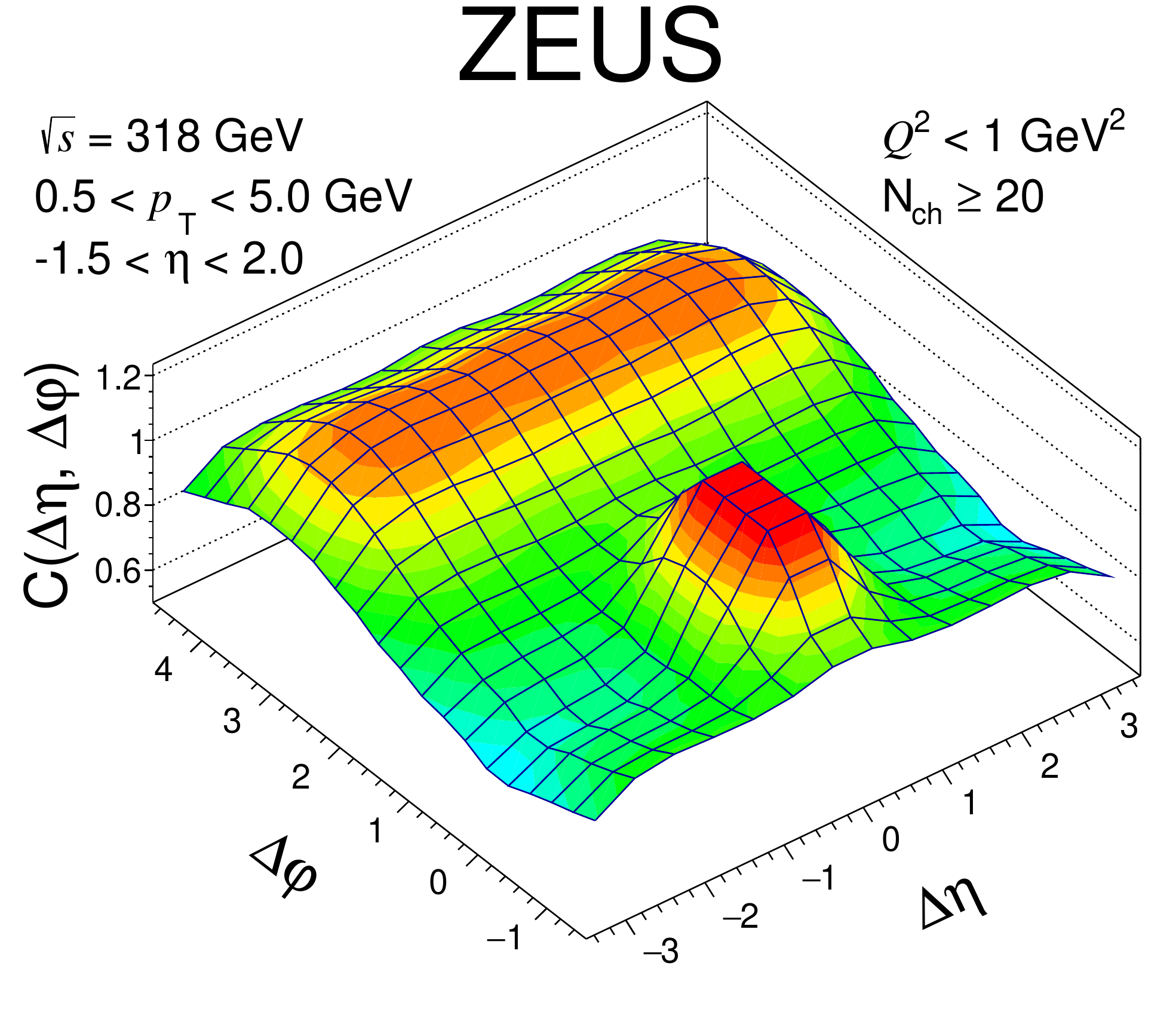}
    \caption{Photoproduction.}
  \label{fig:ridge_php}
\end{subfigure}
\begin{subfigure}[b]{0.49\textwidth}
  \includegraphics[width=\textwidth]{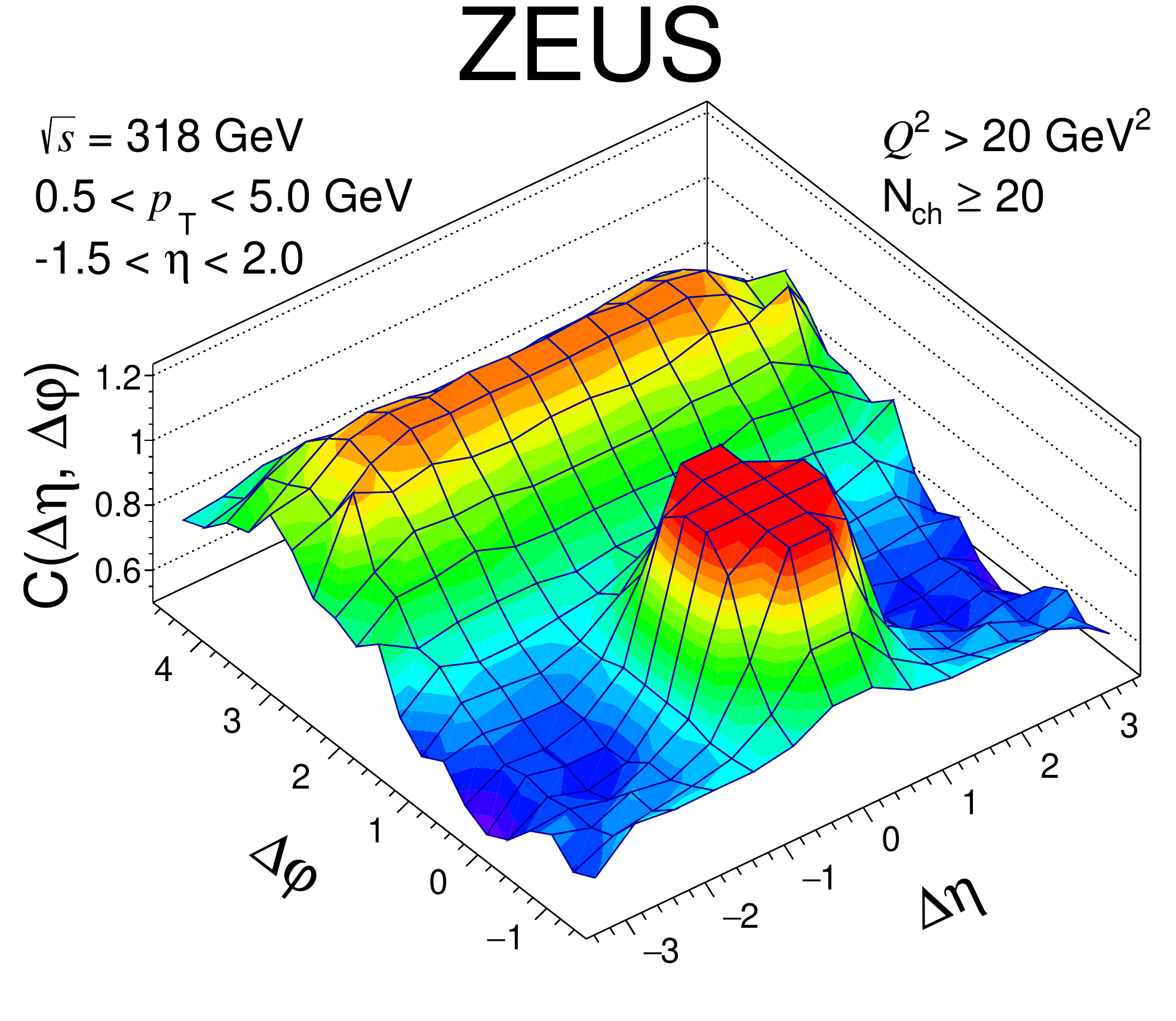}
  \caption{NC DIS with $Q^2 > 20 \GeV^{\;2}$.}
    \label{fig:ridge_dis}
    \end{subfigure}
    \caption{Two-particle correlation \Cdetadphi in (a) photoproduction and (b) NC DIS with $Q^2 > 20 \GeV^{\;2}$.  The peak near the origin has been truncated for better visibility of the finer structures of the correlation.  The plot has been symmetrised along $\Delta\eta$.  No statistical or systematic uncertainties are shown.}
\end{figure}

Figures \ref{fig:ridge_php} and \ref{fig:ridge_dis} show \Cdetadphi in photoproduction and NC DIS, respectively, for particles with $0.5 < \pT < 5.0 \GeV$.
A dominant near-side ($\Delta \varphi$ near 0) peak is seen at small $\Delta \eta$ and $\Delta \varphi$.
On the away side ($\Delta \varphi$ near $\pi$), a broad ridge is observed.
The peak and ridge structures are less pronounced in photoproduction than in NC DIS.
The correlation strength drops for $\deta$ near 3.5 but is consistent with that at $\deta=3.0$ within systematic uncertainties, which are not shown.
As in the previous ZEUS results \cite{ZEUS:2019jya}, there is no indication of a double ridge, which was observed in high-multiplicity $p+p$ and $p$+Pb collisions \cite{Khachatryan:2010gv, Aad:2012gla, Abelev:2012ola}.

\begin{figure}[!ht]
  \centering
  \begin{subfigure}[b]{0.49\textwidth}
  \includegraphics[width=\textwidth]{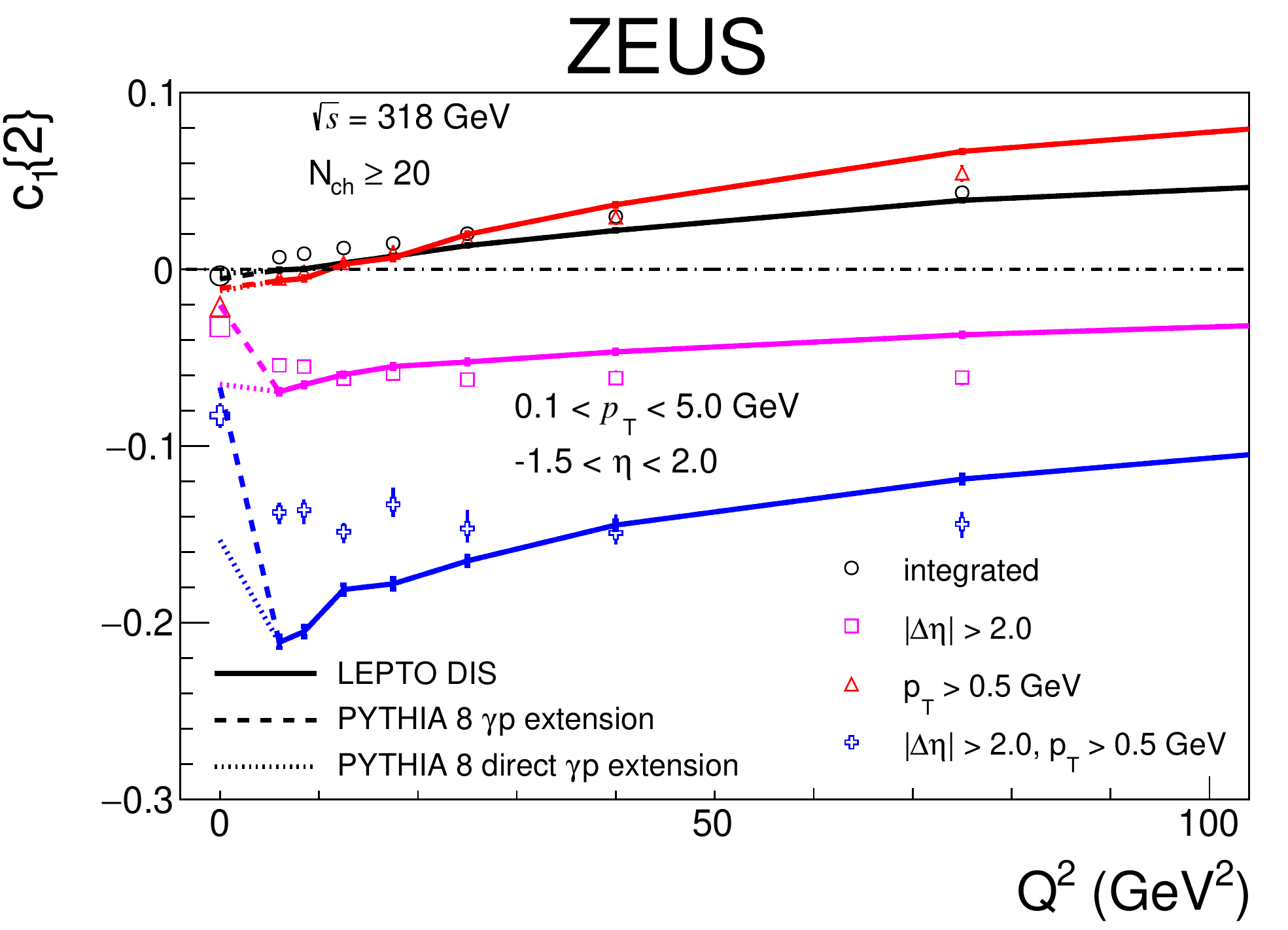}
    \caption{\cn{1} versus $Q^2$.}
  \label{fig:c12_Q2}
\end{subfigure}
\begin{subfigure}[b]{0.49\textwidth}
  \includegraphics[width=\textwidth]{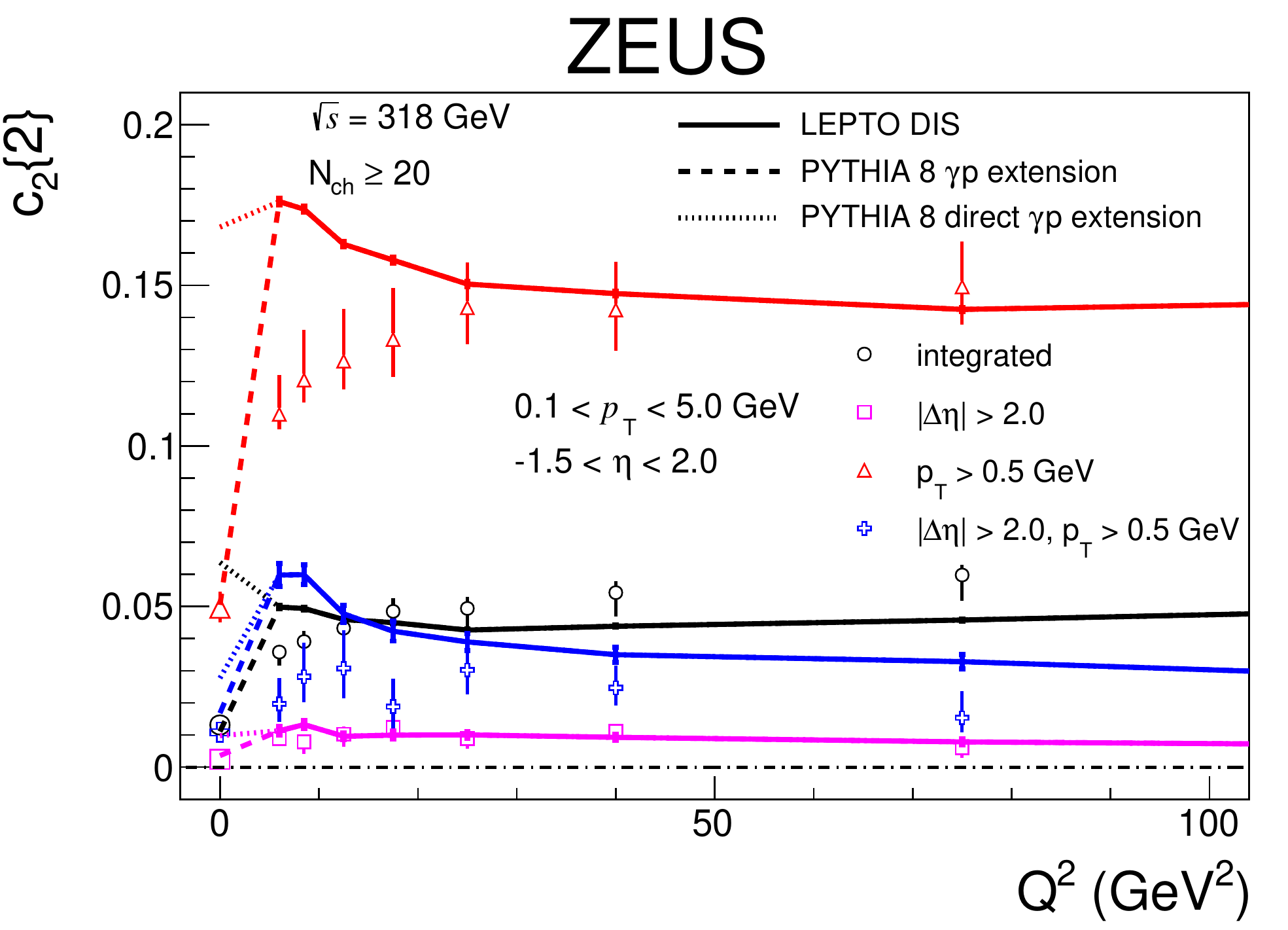}
  \caption{\cn{2} versus $Q^2$.}
  \label{fig:c22_Q2}
  \end{subfigure}
  \caption{Two-particle correlations (a) \cn{1} and (b) \cn{2} versus $Q^2$ with and without a rapidity separation ($\deta > 2$), and for intervals with and without a high-\pT constraint ($\pT>0.5 \GeV$).  Photoproduction data ($\gamma p$) are shown at $Q^2 = 0 \GeV^{\;2}$, while NC DIS is for $Q^2 > 5 \GeV^{\;2}$.  
    LEPTO predictions in NC DIS are shown as solid lines for which the statistical uncertainties are shown as thick vertical lines.  The PYTHIA 8 predictions for $\gamma p$ are shown at the leftmost extension of the dashed lines.  The direct $\gamma p$ contribution alone is similarly shown using a dotted line.  The statistical uncertainties for both PYTHIA predictions are smaller than their respective lines.  The total uncertainties for the ZEUS data are shown as vertical lines.  Zero for \cn{1} is indicated using a dot-dashed line.  Data points are shown at the bin centre.}
\end{figure}

The $Q^2$ dependence of two-particle correlations for the first and second harmonic are shown in Figs.~\ref{fig:c12_Q2} and \ref{fig:c22_Q2}, respectively. 
The results in photoproduction are shown at $Q^2=0 \GeV^2$ and in NC DIS starts at $5 \GeV^2$.
Above $5 \GeV^2$, the $Q^2$ dependence of long-range correlations is observed to be flat.
The magnitude of \cn{1} sharply decreases in photoproduction.
Except for \cn{1} with $\pT>0.5 \GeV$, the correlations in photoproduction are significantly reduced compared to NC DIS.  
The results are presented for the full ranges of \deta and \pT, and with a rapidity-separation condition, $\deta > 2.0$, for $\pT > 0.1$ and $\pT > 0.5 \GeV$.
Short-range correlations that are unrelated to hydrodynamic collective behaviour, such as the near-side peak seen in Figs.~\ref{fig:ridge_php} and \ref{fig:ridge_dis}, are expected to be suppressed by the $\deta > 2.0$ constraint.  
Long-range correlations in heavy-ion collisions are known to increase with \pT up to a few \GeV \cite{Abelev:2013vea, Khachatryan:2010gv, Abelev:2012ola, Aad:2012gla, Aad:2015gqa, Adare:2013piz, Adare:2015ctn}, motivating the additional high-\pT constraint. 
To help further isolate long-range correlations, two-particle correlations with a high-\pT constraint ($\pT > 0.5 \GeV$) are also shown.

\begin{figure}[!ht]
  \centering
  \includegraphics[width=0.49\textwidth]{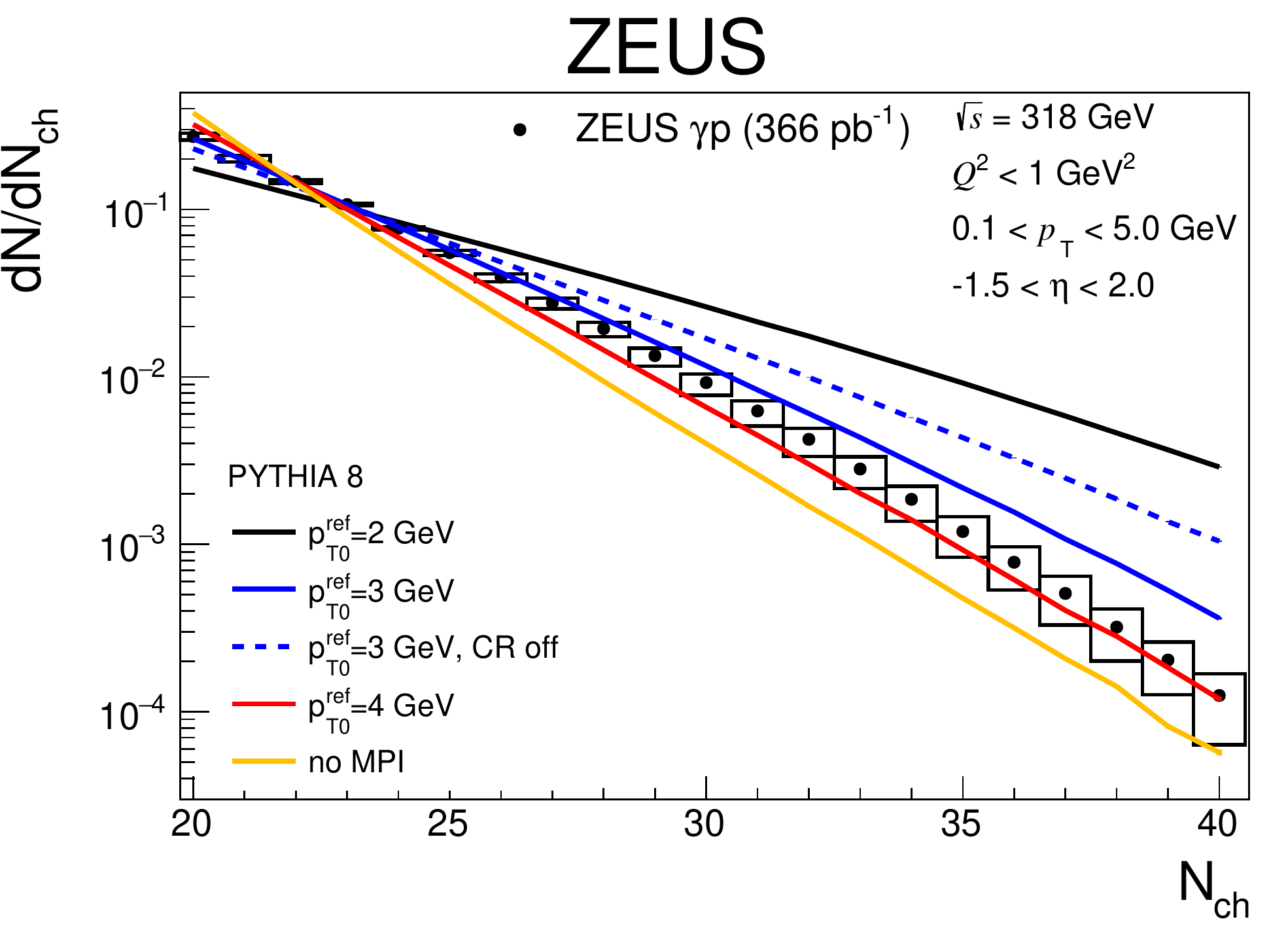}
  \caption{Charged particle multiplicity distribution $dN/d\Nch$ in photoproduction ($\gamma p$) compared to PYTHIA expectations for different levels of MPI, which are inversely related to \pTref. The mean value of the ZEUS distribution is 22.5.  The mean number of MPI for each value of \pTref is: 8.3 ($\pTref=2 \GeV$), 3.8 ($\pTref=3 \GeV$), and 2.2 ($\pTref=4 \GeV$).  The dashed line corresponds to an expectation with colour reconnection (CR) switched off. The integral of the distributions in the range shown are normalised to unity.  The statistical uncertainties are shown as vertical lines although they are typically smaller than the marker size.  Systematic uncertainties are shown as boxes around the data points.  Data points are shown at the bin centre.}
  \label{fig:Nch}
\end{figure}

\begin{figure}[!ht]
  \centering
  \begin{subfigure}[b]{0.49\textwidth}
  \includegraphics[width=\textwidth]{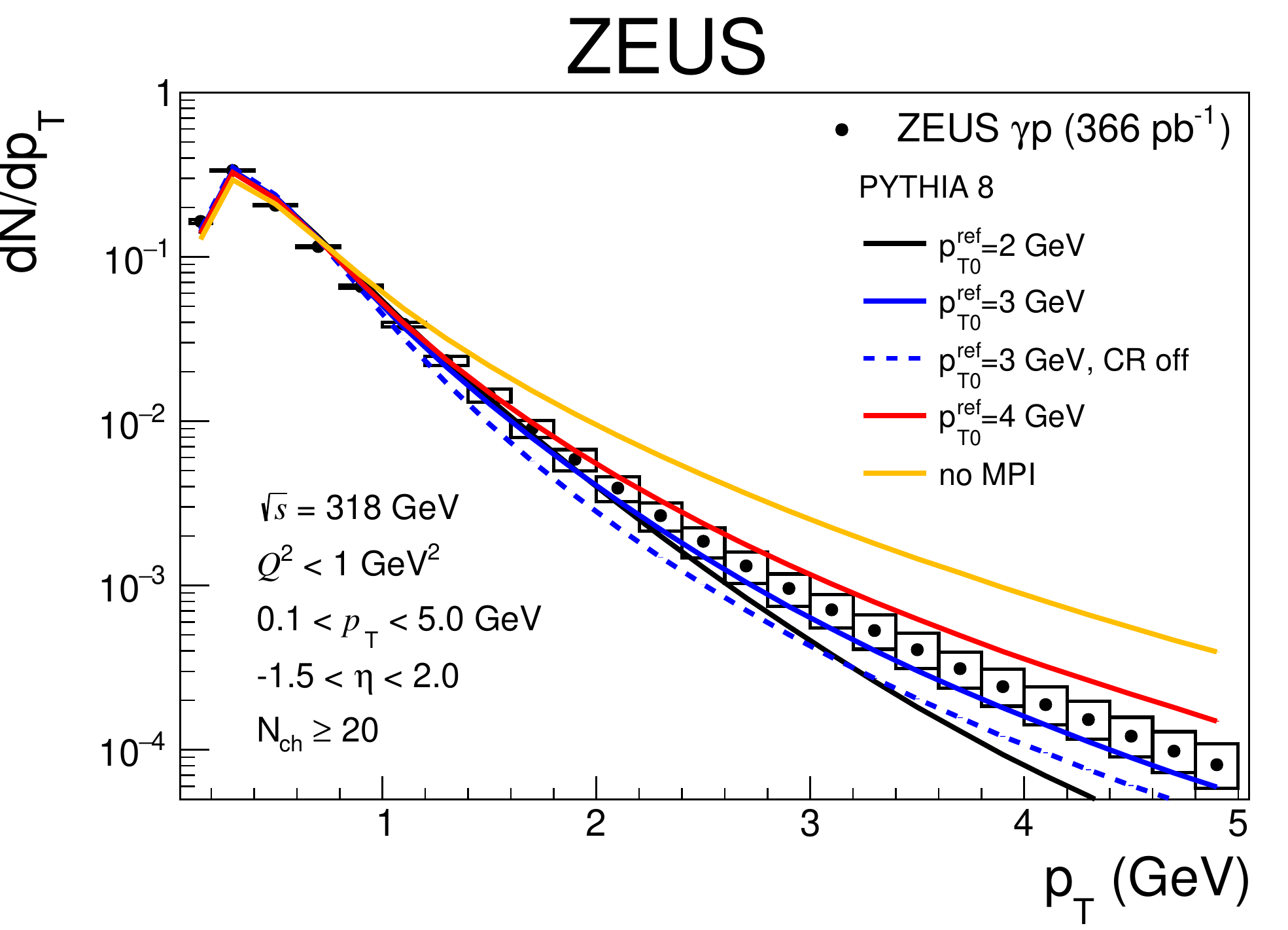}
  \caption{Transverse momentum distribution.}
  \label{fig:pt}
\end{subfigure}
\begin{subfigure}[b]{0.49\textwidth}
  \includegraphics[width=\textwidth]{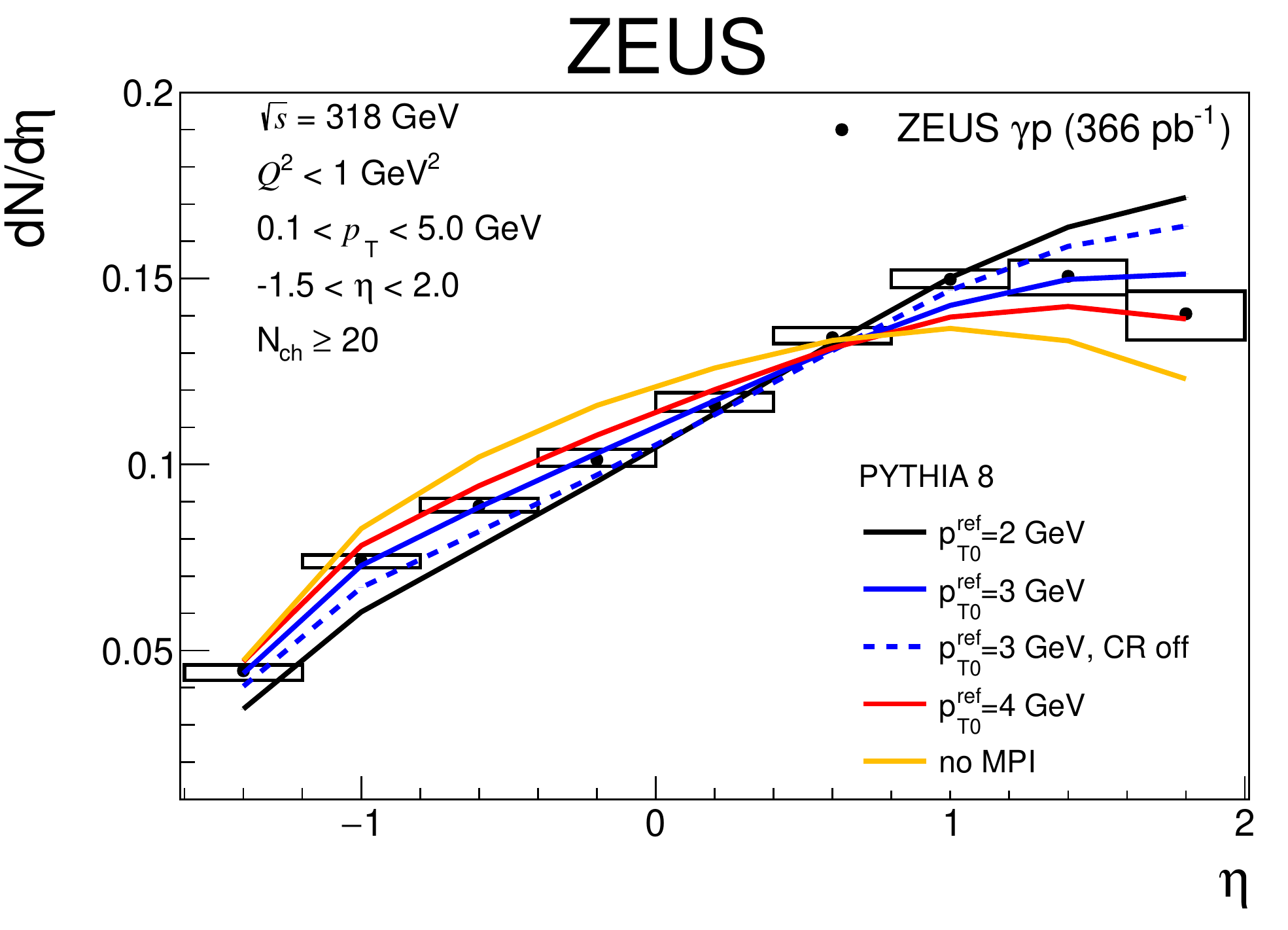}
  \caption{Pseudorapidity distribution.}
  \label{fig:eta}
  \end{subfigure}
  \caption{Normalised charged-particle (a) transverse momentum $dN/d\pT$ and (b) pseudorapidity $dN/d\eta$ distributions compared to PYTHIA expectations for different levels of MPI.  Other details as in Fig.~\ref{fig:Nch}.}
\end{figure}

\begin{figure}[ht]
  \centering
  \begin{subfigure}[b]{0.49\textwidth}
  \includegraphics[width=\textwidth]{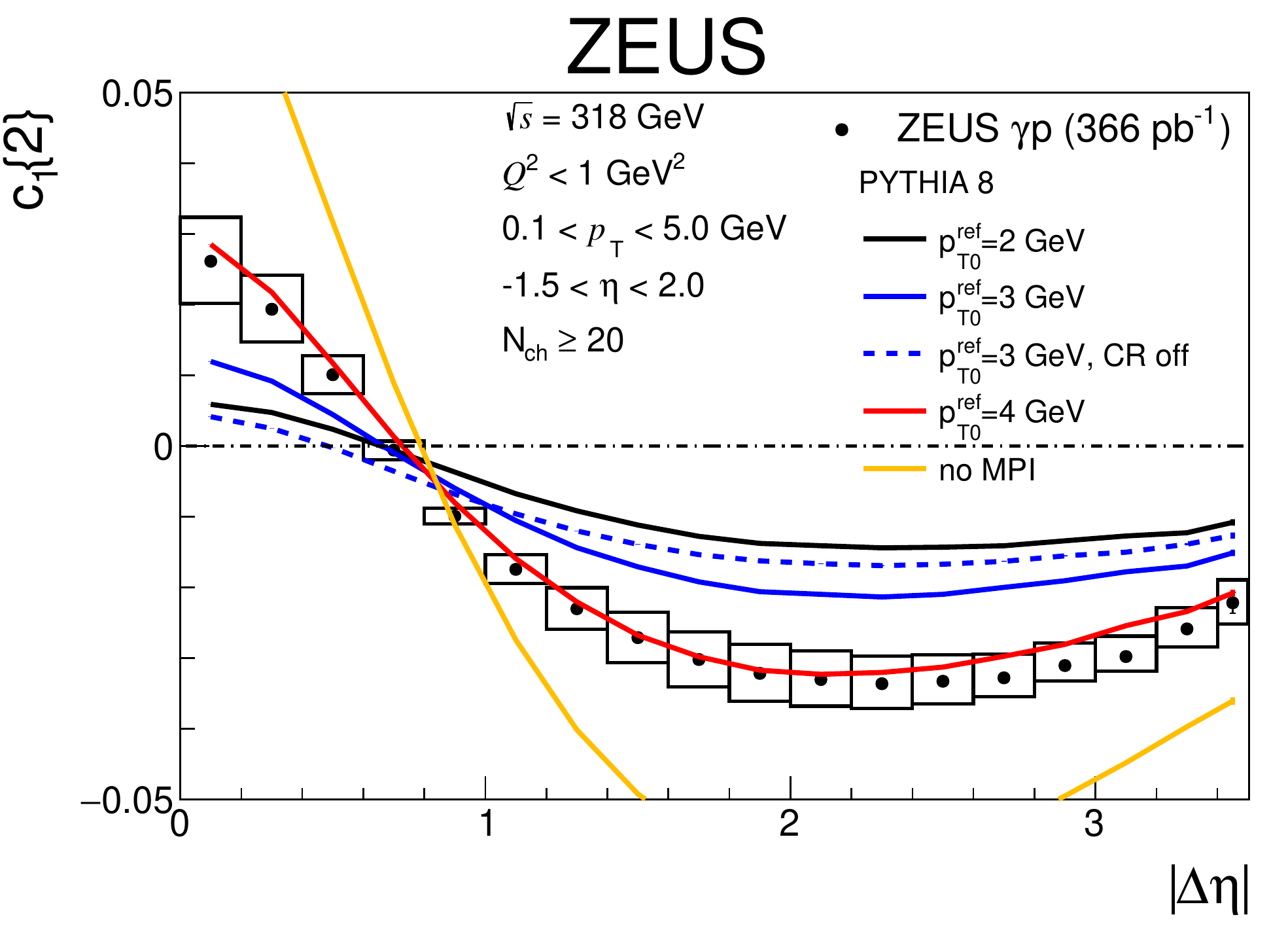}
  \caption{\cn{1} versus \deta.}
  \label{fig:c12_dEta}
  \end{subfigure}
  \begin{subfigure}[b]{0.49\textwidth}
  \includegraphics[width=\textwidth]{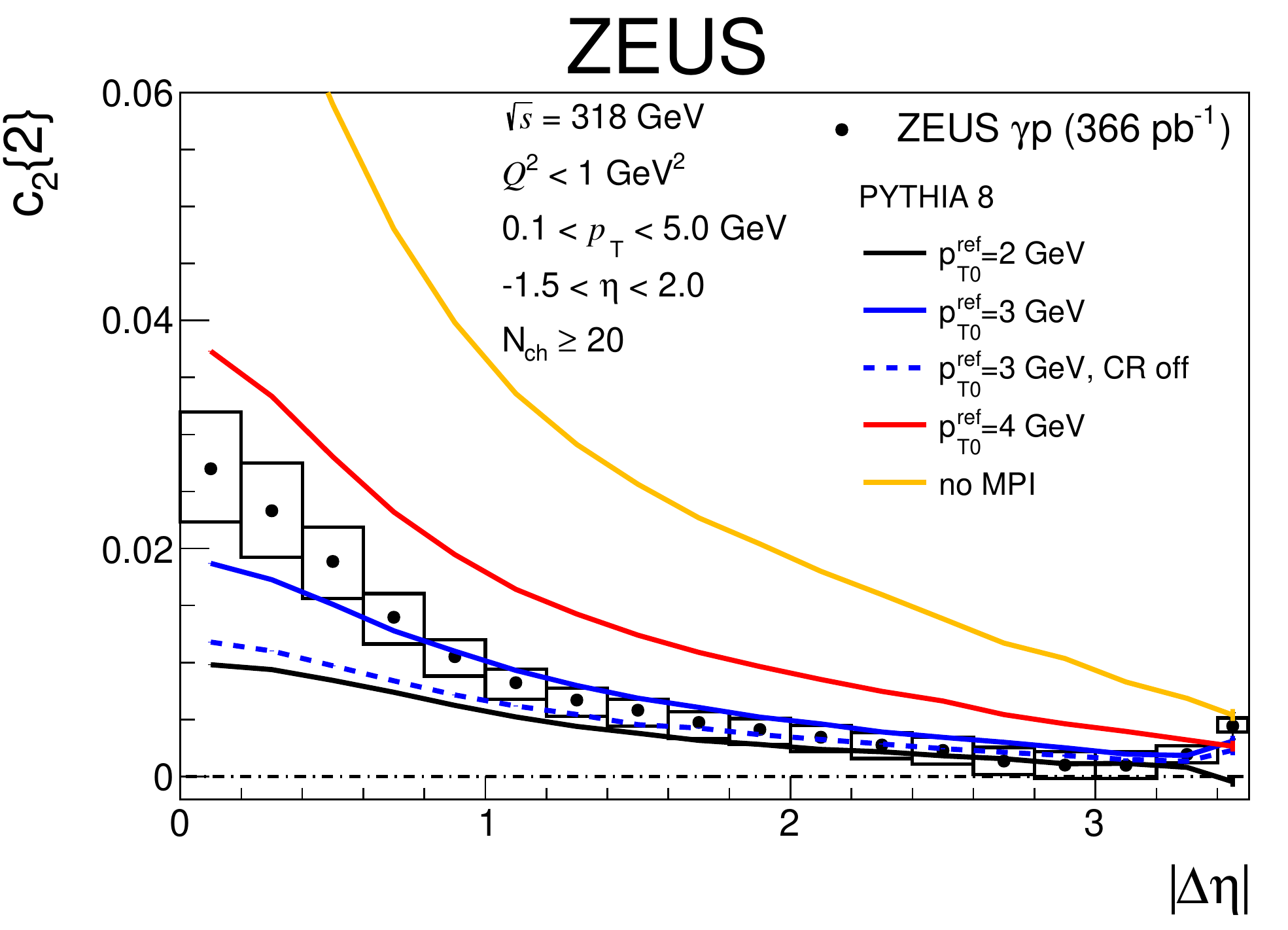}
  \caption{\cn{2} versus \deta.}
  \label{fig:c22_dEta}
  \end{subfigure}
    \caption{Two-particle correlations (a) \cn{1} and (b) \cn{2} versus \deta compared to PYTHIA expectations for different levels of MPI, which are inversely related to \pTref. The mean number of MPI for each value of \pTref is: 8.3 ($\pTref=2 \GeV$), 3.8 ($\pTref=3 \GeV$), and 2.2 ($\pTref=4 \GeV$).  The dashed line corresponds to an expectation with colour reconnection (CR) switched off.  The statistical uncertainties of the data are shown as vertical lines although they are typically smaller than the marker size.  Systematic uncertainties are shown as boxes around the data points.  Statistical errors for the PYTHIA predictions are shown as thick vertical lines.  Data points are shown at the bin centre.}
\end{figure}

\begin{figure}[ht]
  \centering
  \begin{subfigure}[b]{0.49\textwidth}
  \includegraphics[width=\textwidth]{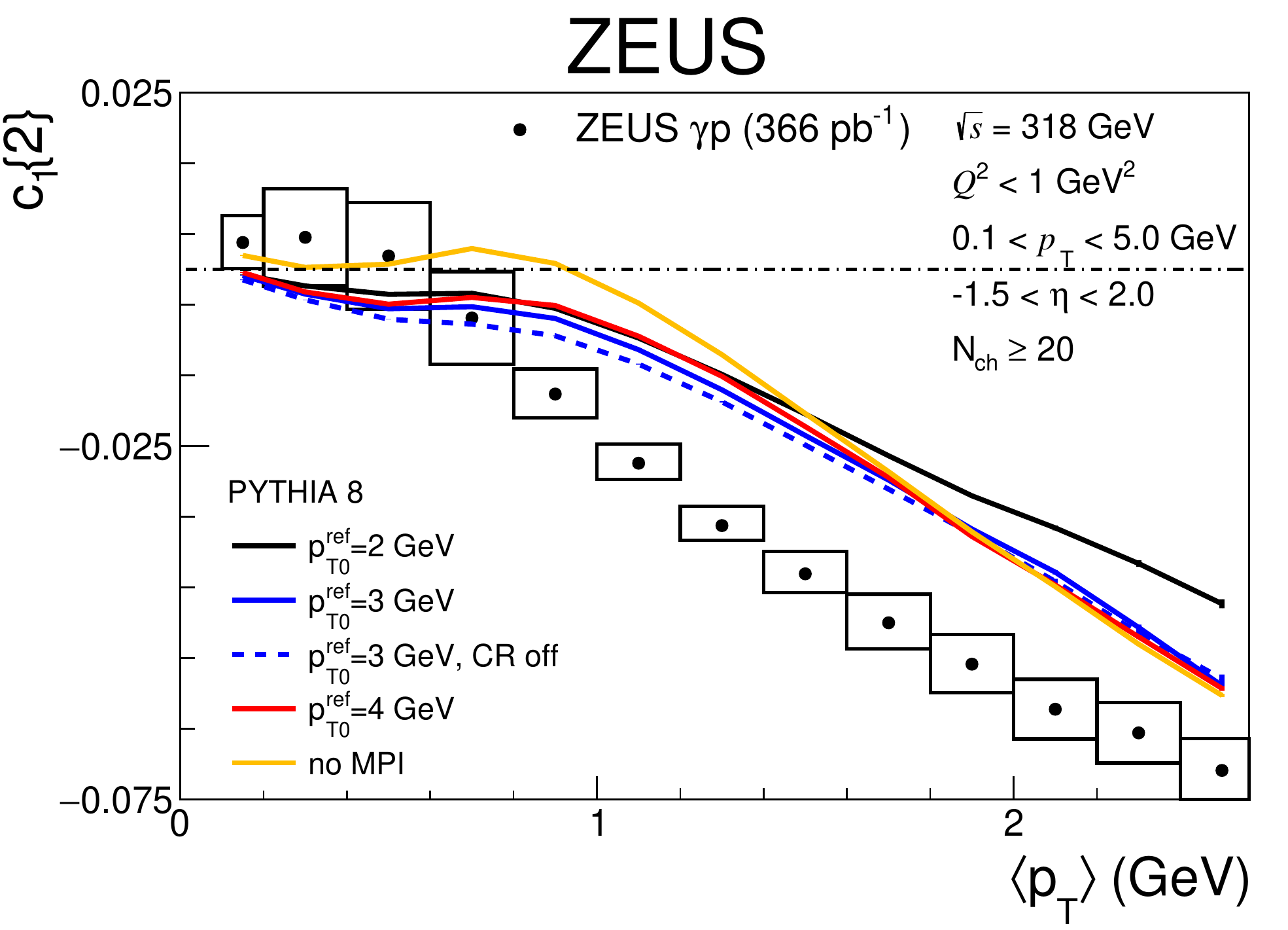}
  \caption{\cn{1} versus \meanpt.}
  \label{fig:c12_mPt}
\end{subfigure}
\begin{subfigure}[b]{0.49\textwidth}
  \includegraphics[width=\textwidth]{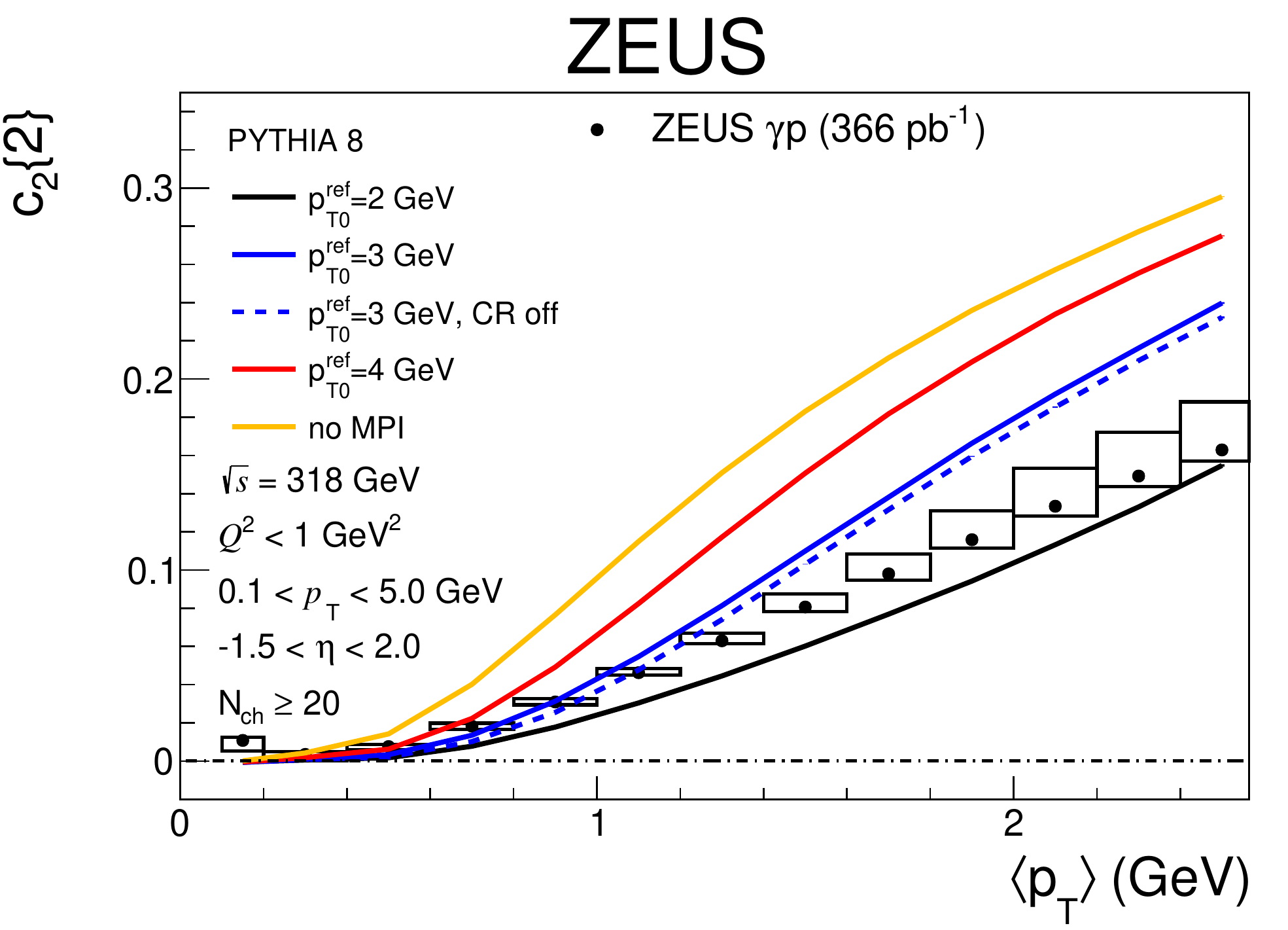}
    \caption{\cn{2} versus \meanpt.}
  \label{fig:c22_mPt}
\end{subfigure}
\caption{Two-particle correlations (a) \cn{1} and (b) \cn{2} versus \meanpt compared to PYTHIA expectations for different levels of MPI.  Other details as in Fig.~\ref{fig:c12_dEta}.}
\end{figure}

\begin{figure}[ht]
  \centering
  \begin{subfigure}[b]{0.49\textwidth}
  \includegraphics[width=\textwidth]{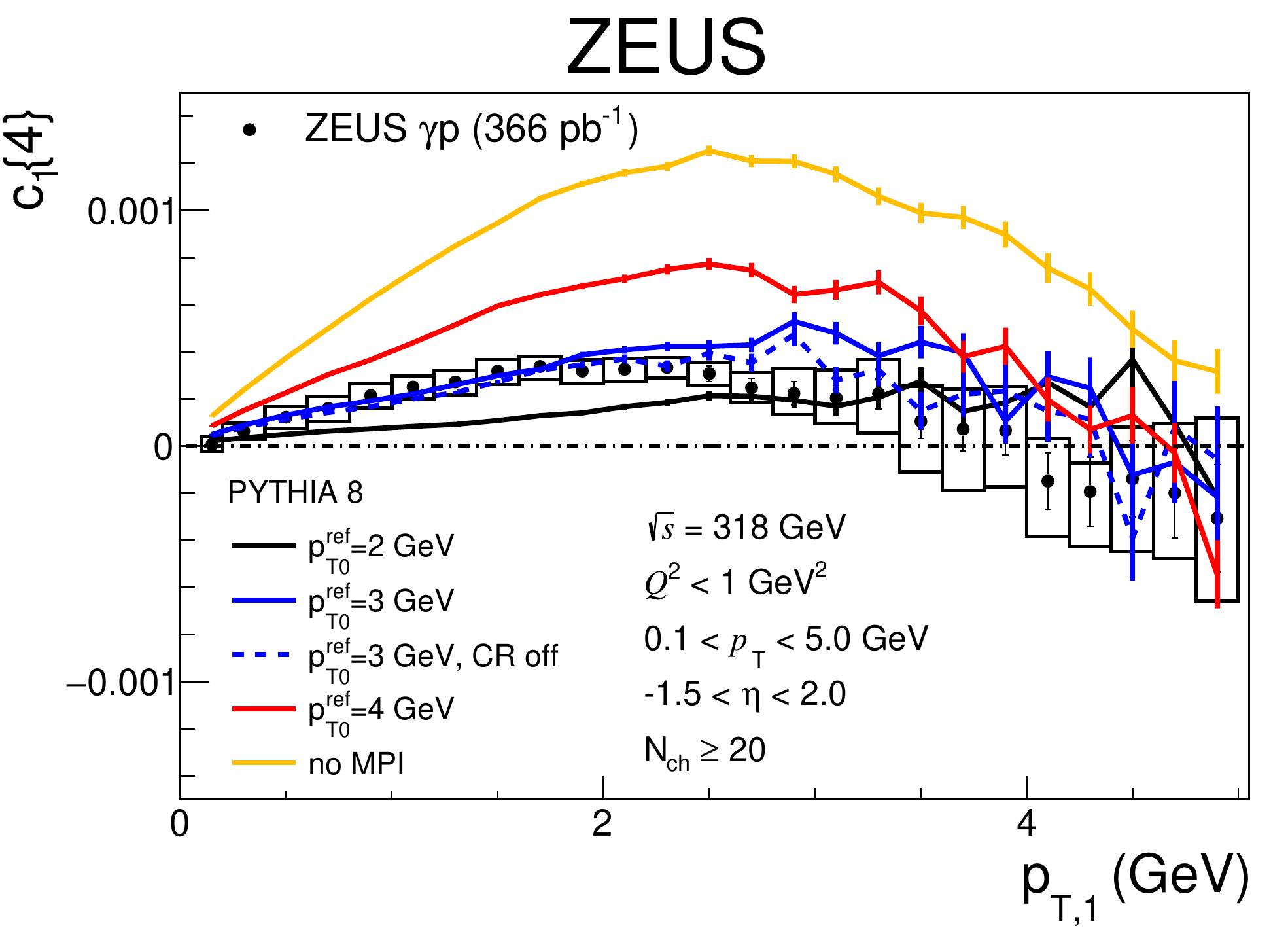}
    \caption{\cnFour{1} versus \pTone.}
    \label{fig:c14_ptpoi}
    \end{subfigure}
    \begin{subfigure}[b]{0.49\textwidth}
    \includegraphics[width=\textwidth]{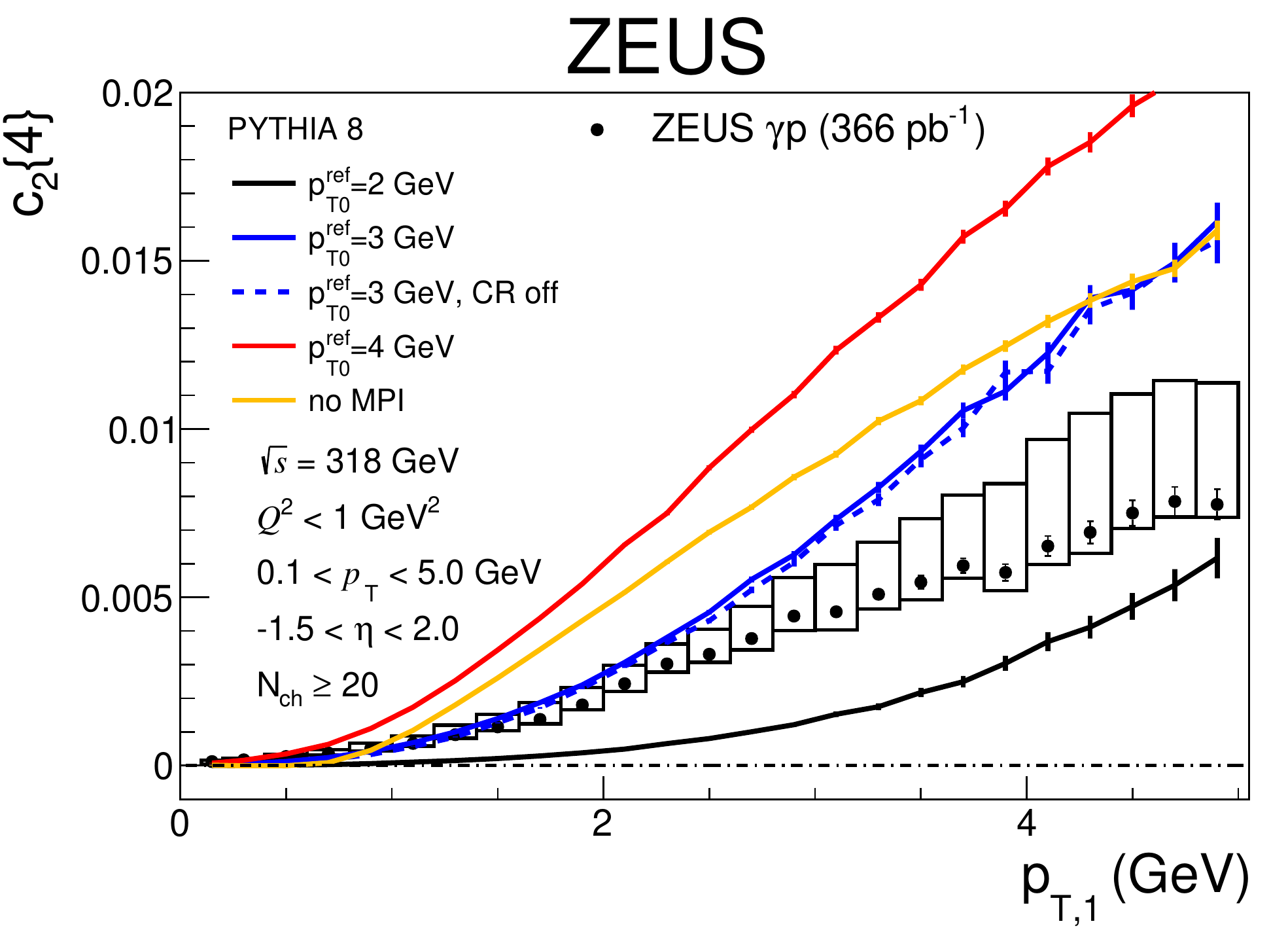}
    \caption{\cnFour{2} versus \pTone.}
    \label{fig:c24_ptpoi}
    \end{subfigure}
    \caption{Four-particle cumulant correlations (a) \cnFour{1} and (b) \cnFour{2} versus \pTone compared to PYTHIA expectations for different levels of MPI.  Other details as in Fig.~\ref{fig:c12_dEta}. }
\end{figure}

The normalised charged-particle multiplicity distribution in photoproduction corrected for tracking inefficiency and the trigger bias is shown in Fig.~\ref{fig:Nch}.
Expectations from PYTHIA are shown with varying levels of MPI and colour reconnection.
The normalised $dN/d\pT$ and $dN/d\eta$ distributions of charged particles in photoproduction are shown in Figs.~\ref{fig:pt} and \ref{fig:eta}, respectively.
Two-particle correlations as a function of \deta for the first and second harmonics are shown in Figs.~\ref{fig:c12_dEta} and \ref{fig:c22_dEta}, respectively.
At low \deta, the correlations are positive and decrease rapidly toward larger \deta.
Two-particle correlations are shown as a function of \meanpt for the first and second harmonics in Figs.~\ref{fig:c12_mPt} and \ref{fig:c22_mPt}, respectively.
For both \cn{1} and \cn{2}, the correlation strength grows with increasing $\meanpt$, a feature which is universally observed in all collision systems \cite{Abelev:2013vea, Khachatryan:2010gv, Abelev:2012ola, Aad:2012gla, Aad:2015gqa, Adare:2013piz, Adare:2015ctn}.
Four-particle cumulant correlations versus the \pT of the first particle, \pTone, are shown in Figs.~\ref{fig:c14_ptpoi} and \ref{fig:c24_ptpoi} for the first and second harmonics, respectively.
Two-particle correlations unrelated to collective behaviour are removed by construction in the four-particle cumulant.
Except for \cnFour{1} at high \pTone, the cumulant correlations are significantly positive, which indicate the presence of genuine four-particle correlations.

\section{Discussion}
\label{Sec:Discussion}

Two-particle correlations in $ep$ photoproduction and NC DIS are markedly different from those observed in high-multiplicity hadronic collisions at RHIC or the LHC. 
Long-range correlations ($\deta > 2$) observed here are large and negative for \cn{1}, while being much smaller and positive for \cn{2}.
In contrast, larger interaction regions produced at RHIC or the LHC show the reverse, where the positive magnitude of \cn{2} is much larger than the negative magnitude of \cn{1} \cite{Aamodt:2011by}.
The same observation is made with \Cdetadphi at large \deta, where the away-side ridge dominates the landscape and no double ridge is visible.

Recent measurements of two-particle correlations in photo-nuclear ultra-peripheral Pb+Pb collisions at the LHC \cite{ATLAS:2021jhn} have also revealed significant long-range correlations.
For a similar kinematic interval ($\left< \Nch \right>$ near 25, $\pT>0.5$, and $\deta>2$), the observed two-particle correlation strengths are consistent with the $ep$ photoproduction values seen in Fig.~\ref{fig:c22_Q2}.
The long-range correlations at such multiplicities are expected via jet production, as seen in Figs.~\ref{fig:c22_dEta} and \ref{fig:c22_mPt} from PYTHIA.
Such correlations are unrelated to hydrodynamic collective behaviour.

PYTHIA 8 predictions ($\pTref=3 \GeV$) at $Q^2 = 0 \GeV^2$ and LEPTO predictions for $Q^2 > 5 \GeV^2$ are compared to two-particle correlations versus $Q^2$ in Figs.~\ref{fig:c12_Q2} and \ref{fig:c22_Q2}.
It is clear that the direct component alone does not describe the photoproduction data well and typically resembles the LEPTO expectations at high $Q^2$.
The inclusion of the resolved component with MPI substantially improves the description of the data.
Comparisons of PYTHIA 6 with MPI to the photoproduction data show good agreement with the data (see the supplementary material).
Between 5 and 20 $\GeV^2$, the data differ significantly from LEPTO, while at higher $Q^2$ the description improves.

The sign of the measured four-particle cumulant correlation differs from measurements in heavy-ion collisions that do not fully overlap, which are negative \cite{Aaboud:2017acw}, as expected from hydrodynamic collective behaviour \cite{Borghini:2001vi}. 
However, this situation is different in small systems such as \ep photoproduction, in which the eccentricity of the initial state as depicted in Fig.~\ref{fig:transverseview_cartoon} fluctuates event-by-event.
In contrast, heavy-ion collisions are characterised by a persistent elliptical eccentricity, which dominates over the event-by-event component induced by fluctuating parton distributions within the nucleus \cite{Acharya:2019vdf}. 

Multiparton interactions in $ep$ photoproduction were investigated through comparisons of $dN/d\Nch$, $dN/d\pT$, $dN/d\eta$, \cn{n}, and \cnFour{n} with PYTHIA predictions.
While there is no consistent preference of the \pTref parameter in PYTHIA in Figs.\ref{fig:Nch}--\ref{fig:c24_ptpoi}, it is clear that the expectation without MPI is never favoured.
The inclusion of MPI in PYTHIA generally increases the number of events at high multiplicity and softens the \pT spectrum.  
Two- and four-particle correlations are most pronounced without MPI and appear diluted by the addition of more independent $2\rightarrow2$ parton scatterings (smaller \pTref) between the photon and proton PDFs.
The more extreme cases of no MPI and high MPI are clearly disfavoured.

\section{Summary and outlook}
\label{Sec:Summary}

Measurements of charged-particle azimuthal correlations have been presented using data taken with the ZEUS detector at HERA in \ep photoproduction and NC DIS at $\sqrt{s}=318 \GeV$ and $\Nch \geq 20$, using an integrated luminosity of $366\pm7\,\mathrm{pb}^{-1}$.
In photoproduction, charged-particle multiplicity, transverse momentum, and pseudorapidity distributions have been presented.

There is no clear indication of a double ridge in \Cdetadphi in either photoproduction or NC DIS at $Q^2 > 20 \GeV^2$.
The evolution of two-particle correlations with $Q^2$ clearly demonstrates that their strength in photoproduction is significantly smaller than in DIS.
The observation indicates the presence of an additional contribution to two-particle correlations beyond that from kinematics associated with the decreasing transverse momentum transferred by the beam electron as $Q^2$ decreases. 
Long-range ($\deta > 2$) correlations observed here with \cn{1} are much more negative than \cn{2} is positive, which is qualitatively different from the kind of the collective behaviour associated with heavy-ion collisions.

PYTHIA 8 expectations have been compared to the photoproduction measurements and the possibility of MPI in \ep scattering has been investigated. 
The comparisons clearly demonstrate sensitivity of the measurements to MPI as parametrised in PYTHIA and provide a strong indication of their presence.
Similar conclusions have been made in previous HERA analyses of jet and particle production \cite{Butterworth:2005aq, Chekanov:2007ab, Newman:2013ada}.
For the PYTHIA predictions with MPI shown in this article, the mean number of distinct $2\rightarrow2$ initial parton scatterings per event in photoproduction ranged from 2 to 8.
Other parameters in PYTHIA such as those pertaining to parton showering and hadronisation are also expected to play an important role and should be investigated.

These measurements provide new insight into the features of azimuthal particle correlations in photon-initiated scattering.
Ongoing measurements at the LHC in ultra-peripheral $A+A$ and $p+A$ collisions as well as future measurements with the Electron Ion Collider will be able to investigate these interesting features further.

\clearpage

\section*{Acknowledgements}
\label{sec-ack}

\Zacknowledge  

We would like to thank Ilkka Helenius for many discussions concerning
photoproduction in PYTHIA as well as Richard Lednicky for discussions of the spatial extent probed in $ep$ scattering.

\vfill\eject


{
\ifzeusbst
  \ifzmcite
     \bibliographystyle{bst/l4z_default3}
  \else
     \bibliographystyle{bst/l4z_default3_nomcite}
  \fi
\fi
\ifzdrftbst
  \ifzmcite
    \bibliographystyle{bst/l4z_draft3}
  \else
    \bibliographystyle{bst/l4z_draft3_nomcite}
  \fi
\fi
\ifzbstepj
  \ifzmcite
    \bibliographystyle{bst/l4z_epj3}
  \else
    \bibliographystyle{bst/l4z_epj3_nomcite}
  \fi
\fi
\ifzbstjhep
  \ifzmcite
    \bibliographystyle{bst/l4z_jhep3}
  \else
    \bibliographystyle{bst/l4z_jhep3_nomcite}
  \fi
\fi
\ifzbstnp
  \ifzmcite
    \bibliographystyle{bst/l4z_np3}
  \else
    \bibliographystyle{bst/l4z_np3_nomcite}
  \fi
\fi
\ifzbstpl
  \ifzmcite
    \bibliographystyle{bst/l4z_pl3}
  \else
    \bibliographystyle{bst/l4z_pl3_nomcite}
  \fi
\fi
{\raggedright
\bibliography{bib/our_references.bib,%
              bib/l4z_zeus.bib,%
              bib/l4z_conferences.bib,%
              bib/l4z_articles.bib,%
              bib/l4z_temporary.bib,%
              bib/l4z_misc.bib}} 
}
\vfill\eject


\newpage

\clearpage

\end{document}